\documentclass[aps,prb,amsmath,amssymb,superscriptaddress,notitlepage,reprint]{revtex4-2}
\usepackage{amsmath}
\usepackage{amssymb}
\usepackage{latexsym}
\usepackage{graphicx}
\usepackage{times,psfrag}
\usepackage{enumerate}
\usepackage{amsmath}
\usepackage{dsfont}
\usepackage{dcolumn}
\usepackage{bm,bbm}
\usepackage{hyperref}
\usepackage{color}
\usepackage{latexsym,amsmath,amssymb,bm,euscript, ulem}
\usepackage{dsfont}
\usepackage{textcomp}
\usepackage{tabularx}
\usepackage{setspace}
\usepackage{sidecap}
\usepackage{placeins}
\usepackage{threeparttable}
\usepackage{multirow}
\usepackage{physics}
\usepackage{tikz-feynman}
\usepackage{pifont}
\allowdisplaybreaks[4]

\let \oldbm \bm
\renewcommand{\vec}[1]{\oldbm{#1}}

\hypersetup{colorlinks=true,citecolor=blue,linkcolor=blue,urlcolor=blue,bookmarksnumbered=true}
\def\bk{{\vec k}}

\def\bA{{\vec A}}
\def\bB{{\vec B}}

\def\bM{{\vec M}}

\def\ba{{\vec a}}
\def\bb{{\vec b}}

\def\bq{{\vec q}}

\def\bO{{\vec O}}

\def\bv{{\vec v}}
\def\bR{{\vec R}}
\def\bG{{\vec G}}

\def\bB{{\vec B}}

\def\br{{\vec x}}
\def\br{{\vec X}}

\def\bm{{\vec m}}

\def\br{{\vec r}}

\def\bxi{{\boldsymbol \xi}}

\newcommand{\beq}{\begin{equation}}
\newcommand{\eeq}{\end{equation}}
\newcommand{\beqarray}{\begin{eqnarray}}
\newcommand{\eeqarray}{\end{eqnarray}}

\begin{document}
	\bibliographystyle{plain}

	\title{Semiclassical dynamics of electrons in space-time crystal: Magnetization, polarization, and current response}%
	
	\author{Qiang Gao}
	\affiliation{Department of Physics, The University of Texas at Austin, TX 78712, USA}

    \author{Qian Niu}
	\affiliation{ICQD/HFNL and School of Physics, University of Science and Technology of China, Hefei, Anhui 230026, China}
	
	\date{\today}
	\begin{abstract}
		A space-time crystal is defined as a quantum mechanical system with both spatial and temporal periodicity. Such a system can be described by the Floquet-Bloch (FB) theory. We first formulate a semiclassical theory by constructing a wave-packet through the superposition of the FB wave functions and derive the equations of motion of FB electrons subjected to slowly varying external fields (not to be confused with the fast-changing Floquet drive), revealing behaviors similar to ordinary Bloch electrons but with quantities modified in the Floquet context. Specifically, we study local magnetic moment due to the self-rotation of the wave-packet, a contribution to total magnetization from the Berry curvature in $\bk$-space, and the polarization of a fully occupied FB band. Based on the semiclassical theory, we can also show the fingerprint of the energy flow in such an energy-non-conserved system. We then discuss the density matrix of a FB system attached to a thermal bath, which allows us to investigate quantities involving many electrons in the non-interacting limit. As an application, we calculate the intrinsic current response in an oblique spacetime metal showing the non-equilibrium nature of the FB system. The current response can also be related to the acoustoelectric effect. Overall, we develop a systematic approach for studying space-time crystals and provide a powerful tool to explore the electronic properties of this exotic system with coupled space and time.
	\end{abstract}

	\maketitle
\section{Introduction}
Periodically driven systems have long been intensively studied given their great controllability and potential to realize many exotic phases of matter or to achieve great performance in various applications, such as Floquet topological insulator~\cite{lindner2011floquet,rechtsman2013photonic,fleury2016floquet}, (space-)time crystal~\cite{li2012space,sacha2015modeling,else2016floquet,xu2018space,gao2021floquet,vzlabys2021six,peng2022topological}, Floquet engineered moir\'e system~\cite{rodriguez2021low}, and even quantum computation~\cite{liu2013floquet,bomantara2018quantum}. The physics behind the periodically driven systems can be well-described by the Floquet(-Bloch) theory~\cite{barone1977floquet,gomez2013floquet,gao2021floquet,gao2022dc}, which is generally non-perturbative and thus is useful for cases with strong field driven~\cite{ho1986floquet,gao2022dc}. Given the strongly driven nature of those systems, there rises a natural question: how does the electron behave or what kind of unique transport property can the periodic drive bring us? This question has been answered by many physicists from different aspects. To put it in simple words, the answer is that the electronic dynamics are largely modified by time variations and so is the electron population (density), which then leads to various unique transport phenomena: shift current~\cite{gao2022dc,morimoto2016topological}, light-induced Hall effect~\cite{dehghani2015out,sato2019microscopic}, and acoustic (spin)-Hall effect~\cite{kawada2021acoustic,mahfouzi2022elastodynamically,zhao2022acoustically}.

Among various periodically driven systems, those having periodic structures in spatial directions, also known as Floquet-Bloch systems or space-time crystals, have attracted more attention as more degrees of freedom being introduced to their Bloch counterpart~\cite{delplace2013merging,berdakin2021spin}. Depending on how the temporal periodicity is related to the spatial periodicity, the space-time crystals can be roughly classified into two categories: rectangular and oblique spacetime crystals~\cite{xu2018space}. The former can often be seen in systems driven by light~\cite{gao2022dc,morimoto2016topological,dehghani2015out,sato2019microscopic} where the periodicity is either purely in time or purely in space, while the latter can be realized as Bloch systems driven by phonons~\cite{gao2021floquet,peng2022topological} that have non-negligible momentum leading to a purely temporal periodicity and other periodicities tilted in space-time (see Refs.~\cite{kawada2021acoustic,mahfouzi2022elastodynamically,zhao2022acoustically} for recent advances in the phonon-driven Bloch systems). For better understanding and further exploring the unique physics in those systems, it becomes increasingly important to know the electron dynamics and its responses to external fields such as static electric or magnetic fields (not to be confused with the time-periodic Floquet driving field). The responses can in general be categorized into equilibrium responses (e.g., magnetization and polarization) and non-equilibrium responses (transport phenomena such as charge or spin current responses)~\cite{xiao2010berry}. One could imagine that those fairly familiar concepts in ordinary Bloch systems would become much more different and hard to reach in the context of the Floquet system due to its distinct nature. Efforts have been given to some of those aspects, for example, Floquet transport~\cite{kohler2005driven,genske2015floquet,puviani2017dynamics} and orbital magnetization in Floquet-Bloch system~\cite{topp2022orbital}.

Our goal in this work is to establish a systematic framework for studying the electronic dynamics and responses to external fields in space-time crystals by directly extending the semiclassical wave-packet formalism developed for Bloch systems~\cite{sundaram1999wave,xiao2010berry}. The general philosophy is that given a space-time crystal, either rectangular or oblique, we can construct a wave-packet as the superposition of the Floquet-Bloch wavefunctions and then treat the external slowly-varying fields (such as static electric or magnetic fields) as weak inhomogeneity in space and time which can be felt by the wave-packet. Thus, a set of semiclassical equations of motions can be derived characterizing the electronic dynamics in the space-time crystal. 
Our scope is then extended to discussions of the physical phenomena involving many yet non-interacting electrons. By using density matrix analysis, we are able to see the non-equilibrium nature of the Floquet systems in the current response of an ensemble of electrons. 

In particular, we find that the behaviors of Floquet-Bloch electrons are similar to ordinary Bloch electrons but with quantities modified in the Floquet context, for example, the Berry curvatures and the energy. We are specifically interested in the local magnetic moment caused by the self-rotation of the wave-packet in the presence of an external magnetic field, the magnetization from the Berry curvature, and the polarization of a fully occupied Floquet-Bloch band, which has two contributions: one is the Bloch polarization modified by the Floquet drive; the other is the Floquet polarization purely from the time variations. The theory also shows the fingerprint of the energy flow in such an energy-non-conserved system. We then discuss the density matrix of a Floquet-Bloch system connected to a heat bath, which allows us to investigate quantities involving many electrons in the non-interacting limit. The density matrix is in general non-diagonal and time-dependent which indicates the inapplicability of the semiclassical scheme to the damped systems, while it becomes diagonal and time-independent when the connection (or the damping) is turned off. As an application of our theory, we calculate the intrinsic current response in an oblique spacetime metal showing the non-equilibrium nature of the Floquet-Bloch system, which can also be interpreted as the acoustoelectric effect.

This paper is organized as the following. In Sec.~\ref{FB_Theory}, we start with the introduction to the Floquet-Bloch theory for periodically driven Bloch systems and the two time-scales scheme used throughout our work. The eigen-wavefunctions of the Floquet-Bloch system are discussed together with its orthonormal conditions. Then, based on the eigen-wavefunctions, we construct the wave-packet and derive the equations of motions of Floquet-Bloch electrons subjected to slowly varying external potentials in Sec.~\ref{semiclassical_dynamics} with more detailed discussions about the magnetization and polarization in that section. In particular, we discuss the energy flow between the Floquet drive and the Bloch system. We also discuss the density matrix of a Floquet-Bloch system attached to a thermal bath in Sec.~\ref{densityMatrixAnalysisSection} to investigate the electronic populations given that such a driven system is indeed a non-equilibrium system. This then allows us to calculate quantities involving many electrons in the non-interacting limit. Having the semiclassical theory established, we proceed in Sec.~\ref{caseStudy} to apply it to cases with oblique space-time structures where non-trivial responses are expected. One specific physical quantity we are interested in is the intrinsic DC current response, which shows the non-equilibrium nature of the Floquet-Bloch systems. Finally, several concluding remarks are given in Sec.~\ref{discussion}.

\section{Floquet-Bloch Hamiltonian, Wave-functions, and Their properties}\label{FB_Theory}
\subsection{Space-time lattice and multi-band Floquet-Bloch system}
Here we give some primes on the Floquet-Bloch formalism and related properties that are useful for our discussions. Our starting point is to construct a Floquet system by time-periodically perturbing a Bloch system. When there is no perturbation, we have the corresponding static system which is basically a multi-band Bloch system:
\begin{equation}
    H_B(\boldsymbol{r}) |\varphi^\alpha_{\boldsymbol{k}}(\boldsymbol{r})\rangle = E^\alpha (\boldsymbol{k})|\varphi^\alpha_{\boldsymbol{k}}(\boldsymbol{r})\rangle
\end{equation}
where $H_B(\boldsymbol{r})=H_B(\boldsymbol{r}+\boldsymbol{R})$ is the Hamiltonian for the static Bloch system with $N_B$ multiple bands, $\alpha=\{1,2,\cdots,N_B\}$ stands for the Bloch band index with $\alpha = 1$ being the ground state, and $\boldsymbol{k}$ stands for lattice momentum. From Bloch's Theorem, the wavefunction can be expressed as a plane wave multiplied with a space periodic function: $|\varphi^\alpha_{\boldsymbol{k}}(\boldsymbol{r})\rangle = e^{i\boldsymbol{k}\cdot\boldsymbol{r}}|u^\alpha_{\boldsymbol{k}}(\boldsymbol{r})\rangle$, where $|u^\alpha_{\boldsymbol{k}}(\boldsymbol{r}+\boldsymbol{R})\rangle = |u^\alpha_{\boldsymbol{k}}(\boldsymbol{r})\rangle$.

\begin{figure}[t]
    \centering
    \includegraphics[width=0.49\textwidth]{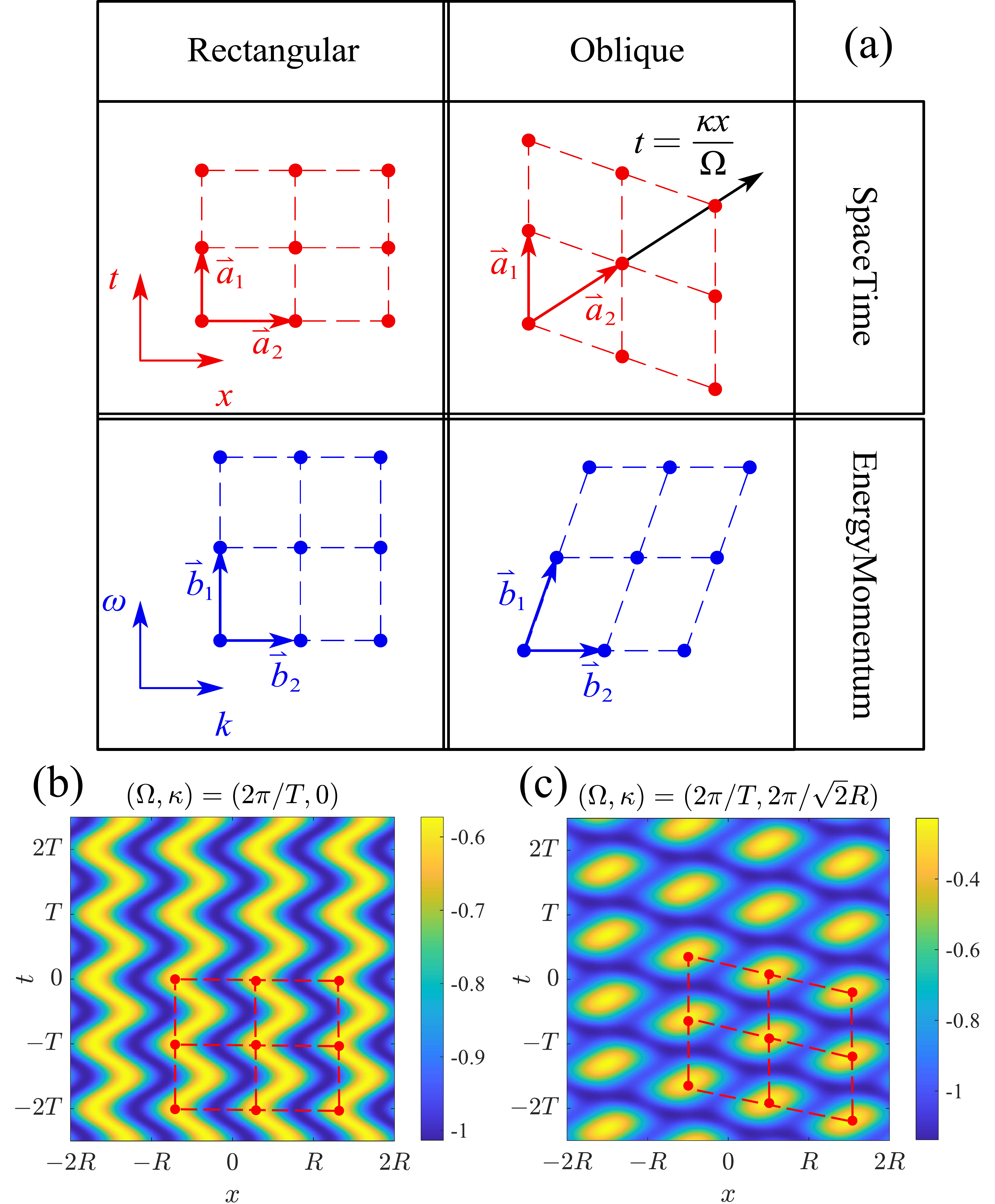}
    \caption{(a) The illustration of two types of simple spacetime lattices in (1+1)D: the rectangular spacetime lattice (left two panels) and the oblique spacetime lattice (right two panels) in both real ($t,x$) space (top two panels) and reciprocal ($\omega,k$) space (bottom two panels). (b,c) The space-time potential following Eq.~\eqref{spacetimepotential} with $V_0 =1$, $C=0.2R$, and $\sigma = 0.2R^2$ for rectangular (b) and oblique (c) space-time crystals. The red dots highlight the underline space-time lattice structures.}
    \label{spacetimeLattice}
\end{figure}

Now, we can introduce the time-periodic perturbation to the Bloch system which makes $H_B \rightarrow H(\boldsymbol{r},t)$ with periodicity in time $H(\boldsymbol{r},t) = H(\boldsymbol{r},t+T)$, and the Schrodinger equation becomes time-dependent:
\begin{equation}\label{timeSE}
    H(\boldsymbol{r},t)|\Psi(\boldsymbol{r},t)\rangle = i\hbar\partial_t |\Psi(\boldsymbol{r},t)\rangle.
\end{equation}
Normally, the new Hamiltonian should inherit the spatial periodicity from the static Bloch Hamiltonian that $H(\boldsymbol{r},t) = H(\boldsymbol{r}+\boldsymbol{R},t)$. However, in this note we are going to discuss a more general scenario where the spatial periodicity can be partly broken by the time variation ending up with a coupled periodicity in both temporal and spatial direction: $H(\boldsymbol{r},t) = H(\boldsymbol{r}+\boldsymbol{R},t+\frac{\boldsymbol{\kappa}\cdot R}{\Omega})$ where $\Omega = 2\pi/T$ and $\boldsymbol{\kappa}$ are the energy and momentum introduced by the time variation. We call a system with a zero $\boldsymbol{\kappa}$ the rectangular space-time crystal and a system with a non-zero $\boldsymbol{\kappa}$ the oblique space-time crystal. We want to emphasize that, for systems considered in this work, the temporal periodicity $ H(\br,t+T)=H(\br,t)$ is exact regardless of $\kappa$ being zero or nonzero.
In Fig.~\ref{spacetimeLattice} (a), we show those two types of spacetime lattices in both the real space (space-time) and the reciprocal space (energy-momentum). We note here that the spacetime we used adopts a signature $(-_{t,\omega},+_{x,k})$, so that $\ba_i\cdot\bb_j = \pm 2\pi\delta_{ij}$ holds. 

Then, as an illustration, we consider a simple $(1+1)$D Hamiltonian $H(x,t)=\frac{-\hbar^2}{2m}\partial_{x}^2+V(x,t)$ with a space-time potential
\begin{equation}\label{spacetimepotential}
    V(x,t) = -V_0\sum_{l}e^{-(x-x_l(t))^2/\sigma},
\end{equation}
where $V_0$ is the potential constant, $\sigma$ controls the width of the potential, and
$
    x_l(t) \equiv lR - C\cos(\kappa lR-\Omega t)
$
is the position of $l^{\text{th}}$ ion. This specific ion position profile describes a sound wave with mode $(\Omega,\kappa)$ and amplitude $C$ propagating through the 1D lattice (which serves as the fast-varying Floquet driving field). We note that setting $C=0$ reduces the system to its corresponding static Bloch system. We can check that the space-time potential satisfies the correct periodicity: $V(x,t+T)=V(x,t)$ and $V(x+R,t+\frac{\kappa R}{\Omega})=V(x,t)$. We plot the space-time potential for two special cases: $(\Omega,\kappa)=(2\pi/T,0)$ corresponding to a rectangular space-time lattice [Fig.~\ref{spacetimeLattice} (b)] and $(\Omega,\kappa)=(2\pi/T,2\pi/\sqrt{2}R)$~\cite{Note_kappa} corresponding to an oblique space-time lattice [Fig.~\ref{spacetimeLattice} (c)].

Now, we have gained some basic knowledge about the space-time crystalline structure. Let us then move to the discussion about its wavefunctions. Analogous to the Bloch crystal where the Bloch wavefunctions are eigenstates of both the Hamiltonian and the translation operators, here we should have the Floquet-Bloch wavefunction to be the eigenstate of both the Floquet-Bloch Hamiltonian and its corresponding translation operators. Thus, for a Floquet-Bloch eigenstate labeled by energy-momentum $\bxi\equiv(\omega,\bk)$, the following equations should be satisfied:
\begin{equation}
    \begin{split}
        &\hat{T}_{\ba_1}|\Psi_{\bxi}(\br,t)\rangle = |\Psi_{\bxi}(\br,t+T)\rangle\\
        &\qquad= e^{i\ba_1\cdot\bxi}|\Psi_{\bxi}(\br,t)\rangle = e^{-i\omega T}|\Psi_{\bxi}(\br,t)\rangle, \\
        &\hat{T}_{\ba_2}|\Psi_{\bxi}(\br,t)\rangle = |\Psi_{\bxi}(\br+\bR,t+\frac{\boldsymbol{\kappa}\cdot\bR}{\Omega})\rangle\\
        &\qquad= e^{i\ba_2\cdot\bxi}|\Psi_{\bxi}(\br,t)\rangle = e^{-i\omega \frac{\boldsymbol{\kappa}\cdot\bR}{\Omega}+i\bk\cdot\bR}|\Psi_{\bxi}(\br,t)\rangle,
    \end{split}
\end{equation}
where $\hat{T}_{\ba_1} $ and $\hat{T}_{\ba_2}$ are two translation operators along two translation symmetric directions: $\ba_1 = (T,0)$ and $ \ba_2 = (\frac{\boldsymbol{\kappa}\cdot\bR}{\Omega},\bR)$ [also shown in Fig.~\ref{spacetimeLattice}(a)]. We can construct a set of basis wavefunctions~\cite{gao2021floquet,peng2022topological} that satisfy the above equations, which read
\begin{equation}
    |\Psi^{\alpha}_{n,\bxi}(\br,t)\rangle = e^{-i\omega t}|\Phi^\alpha_{n,\boldsymbol{k}}(\boldsymbol{r},t)\rangle \equiv e^{-i(\omega + n\Omega) t}|\varphi^\alpha_{\boldsymbol{k}+n\boldsymbol{\kappa}}(\boldsymbol{r})\rangle,
\end{equation}
which is nothing but the original Bloch wavefunction shifted both in energy and momentum by an integer multiple of $(\Omega,\boldsymbol{\kappa})$. Here $n$ is the so-called Floquet band index. Thus, true eigenstates can be expressed as linear combinations of those basis wavefunctions
\begin{align}\label{FBeigenstate}
    &|\Psi^\mu_\bk(\br,t)\rangle=\sum_{n,\alpha}f^{\mu,\alpha}_{n,\boldsymbol{k}}|\Psi^{\alpha}_{n,\bxi}(\br,t)\rangle \notag\\
    &= e^{-i\omega_\mu(\boldsymbol{k})t}\sum_{n,\alpha}f^{\mu,\alpha}_{n,\boldsymbol{k}}|\Phi^\alpha_{n,\boldsymbol{k}}(\boldsymbol{r},t)\rangle\equiv e^{-i\omega_\mu(\boldsymbol{k})t}e^{i\boldsymbol{k}\cdot\boldsymbol{r}}|\Tilde{u}_{\boldsymbol{k}}^\mu\rangle
\end{align}
where we have added the Floquet band index $\mu$ and denoted the quasi-energy $\omega_\mu$, $|\Tilde{u}_{\boldsymbol{k}}^\mu\rangle$ refers to the Floquet-Bloch periodic function that has the same periodicity as the Hamiltonian, and the coefficient $f^{\mu,\alpha}_{n,\boldsymbol{k}}$ satisfies the following eigen-equation:$\sum_{n,\beta}H_{m,n;\alpha,\beta}(\boldsymbol{k})f^{\mu,\beta}_{n,\boldsymbol{k}}=\hbar\omega_\mu(\boldsymbol{k})f^{\mu,\alpha}_{n,\boldsymbol{k}}$ with the kernel matrix defined as
\begin{equation}\label{HamiltonianKernel}
    \begin{split}
    &H_{m,n;\alpha,\beta}(\boldsymbol{k}) \equiv \langle\langle \Phi^\alpha_{m,\boldsymbol{k}}(\boldsymbol{r},t) | H(\boldsymbol{r},t)-i\hbar\partial_t|\Phi^\beta_{n,\boldsymbol{k}}(\boldsymbol{r},t)\rangle\rangle \\
    \end{split}
\end{equation}
There are a few things that need to be clarified: the inner product now also takes the time dimension into account which is defined as $\langle\langle\cdot\rangle\rangle \equiv 1/T\int_0^T \langle\cdot\rangle dt$; the quasi-energy $\omega_\mu$ inherits the Bloch band index so that in principle, we will have $N_B$ different quasi-energy bands $\{\omega_1,\omega_2\cdots\omega_{N_B}\}$ within a Floquet-Bloch zone; the wavevector $\boldsymbol{k}$ is now only conserved up to a change of integer multiples of $\boldsymbol{\kappa}$ so that we can refer it as a quasi-momentum. The kernel matrix $H_{m,n;\alpha,\beta}$ is actually a direct product of the Bloch system and the Floquet degrees of freedom.

In this work, we will keep using the following convention for notions:
\begin{equation}
    \begin{split}
        &\{ \mu,\nu \} \text{ label Floquet bands} \\
        &\{ \alpha,\beta \} \text{ label original Bloch bands}
        \\
        &\{ n,m,l,\cdots \} \text{ label Floquet band replicas}
    \end{split}
\end{equation}
For more than two Floquet (Bloch) bands involved, we use primed symbols to distinguish different states: $\mu,\nu,\mu',\nu',\cdots$ ($\alpha,\beta,\alpha',\beta',\cdots$). Additionally, the periodic part of the Floquet-Bloch wavefunction will also have a tilde on top of it: $|\Tilde{u}_{\boldsymbol{k}}^\mu\rangle$, as compared to that of the Bloch wavefunction that has no tilde: $|u^\alpha_{\boldsymbol{k}}\rangle$.

One of the reasons to choose the extended Floquet-Bloch basis is that we can utilize some properties of the corresponding Bloch system directly. For example, one can obtain the following orthonormal conditions for the wavefunction and constructing basis:
\begin{equation}\label{usefulConditions}
    \begin{split}
        \langle \Psi^\mu_{\boldsymbol{k}}(\boldsymbol{r},t)|\Psi^\nu_{\boldsymbol{k}'}(\boldsymbol{r},t)\rangle = \sum_{\bG}\delta_{\mu,\nu} \delta(\boldsymbol{k}-\boldsymbol{k}'+\bG)
    \end{split}
\end{equation}
where $\bG$ is the reciprocal vector of the corresponding static Bloch lattice.
We note that the above relation is true given the following relation (see Appendix~\ref{orthonormalCondition}):
\begin{equation}\label{orthogonality}
    \sum_{n,\beta}(f_{n+m,\boldsymbol{k}-m\boldsymbol\kappa}^{\mu ,\beta})^*f_{n,\bk+\bG}^{\nu,\beta}=\delta _{\mu ,\nu}\delta_{m,0},
\end{equation}
which follows from the orthogonality of the eigenvector of the kernel matrix in Eq.~\eqref{HamiltonianKernel}. We also utilized the fact that the coefficients also inherit the Bloch periodicity: $f_{n,\bk+\bG}^{\nu,\alpha} = f_{n,\bk}^{\nu,\alpha}$~\cite{Note1}. In the remaining of this work, we shall restrict the quasi-momentum $\bk$ in our discussion to the first Brillouin zone of the corresponding static Bloch lattice.
Then, we have the following completeness relation for the Floquet-Bloch states:
\begin{align}
    \mathcal{I} =& \sum_{\mu} \int d\bk|\Psi^\mu_{\boldsymbol{k}}(\boldsymbol{r},t)\rangle\langle \Psi^\mu_{\boldsymbol{k}}(\boldsymbol{r},t)| \notag\\
    =& \sum_{\mu}\sum_{\alpha\beta,mn}\int d\bk|\Phi^{\alpha}_{m,\bk}(\br)\rangle f^{\mu,\alpha}_{m,\bk} \left( f^{\mu,\beta}_{n,\bk} \right)^*\langle  \Phi^{\beta}_{n,\bk}(\br)|,
\end{align}
or at a specific $\bk$, the completeness relation for the periodic part of the Floquet-Bloch wavefunction:
\begin{equation}
    \mathcal{I}_{\bk} = \sum_\mu |\Tilde{u}_{\boldsymbol{k}}^\mu\rangle\langle \Tilde{u}_{\boldsymbol{k}}^\mu|.
\end{equation}

\subsection{Two Time Scales}
In the last section, we primarily consider Floquet systems with only one fast-varying driving field (light or sound wave) which introduces a typical time scale $T=2\pi/\Omega$. However, in many cases, those systems are also subjected to other external fields with a much slower time scale. Such separation in time scales allows us to treat those two types of driving fields differently: the Floquet Hamiltonian with fast-changing fields as the basis and the slowly-varying fields as perturbations. In this work, we will consider an extended system by adding an external field $\bA$ to the Hamiltonian in Eq.~\eqref{timeSE}, and to further distinguish the two time scales, we explicitly write the extended Hamiltonian as
\begin{equation}
H(\boldsymbol{r},t;\boldsymbol{A}(\boldsymbol{r},t))\equiv H(\boldsymbol{r},\tau_f;\boldsymbol{A}(\boldsymbol{r},\tau_s))
\end{equation}
where the fast time evolution denoted by $\tau_f$ is the Floquet oscillation (for example the space-time potential introduced in Eq.~\eqref{spacetimepotential}, which can be recast as $V(x,\tau_f)$ using the notation defined here) and the slow time evolution denoted by $\tau_s$ is characterized by the external field $\boldsymbol{A}(\boldsymbol{r},\tau_s)$ (such as static electromagnetic fields $\boldsymbol{E}$, $\boldsymbol{B}$) which can be viewed as the envelope function over the fast-changing part.
Although two different labels have been used to distinguish two time scales, they are essentially the same (any time-dependent operator acts on both of them). The two-time-scale method has been adopted for studying the transition behaviors of time periodically driven systems before~\cite{breuer1989quantum,althorpe1997time}.

However, this work focuses on the effect of applying slowly varying external fields. A common technique is to average out the fast oscillating part of the system and what is left is the slow-time evolving part we are interested in. By averaging out the fast-time variations, we end up with the following approximation:
\begin{equation}\label{averageFast}
    \begin{split}
        \langle g(\tau_f,\tau_s)\rangle_T &\approx \Bar{g}(\tau_s)\equiv\frac{1}{T}\int_{\tau_s}^{\tau_s +T} g(\tau_f,\tau_{s0}) d\tau_f \\
        \langle\partial_{\tau_s} g(\tau_f,\tau_s)\rangle_T &\approx \partial_{\tau_s} \Bar{g}(\tau_s); \\
        \langle\partial_{t} g(\tau_f,\tau_s)\rangle_T &\approx \langle\partial_{\tau_f} g(\tau_f,\tau_{s0})\rangle_T \\
        \int dt g(\tau_f,\tau_s) &\approx \int d\tau_s \Bar{g}(\tau_s)
    \end{split}
\end{equation}
where $g(\tau_f,\tau_s)$ is an arbitrary function having two time scales as described before, $\langle\cdot\rangle_T\equiv \frac{1}{T} \int_{\tau_s}^{\tau_s+T}dt$ is to average over a Floquet period. One should notice that the only approximation made in the above equations is that we keep the slowly varying part unchanged (by setting $\tau_s = \tau_{s0} \in [\tau_s, \tau_s+T]$) during the time integration, which is the essence of slow evolution.

We have to emphasize that the main technique used in this work is that we separate the time scale into fast and slowly changing parts and assign two different and independent time variables ($\tau_f$ and $\tau_s$) to describe those two evolutions. The independence between $t$ and $\tau_s$ here is really an assumption that when we integrate out or differentiate the fast time variable $t$, those quantities slowly depending on time characterized by $\tau_s$ are thought to be unchanged. In the cases where we need to count all time variations, we shall denote the total time derivative as
\begin{equation}
    \frac{d}{d t}= \frac{d}{d \tau_f} + \frac{d}{d \tau_s}.
\end{equation}

\section{Wave-packet dynamics for Floquet-Bloch electrons} \label{semiclassical_dynamics}
\subsection{Construction of wave-packet}
In this section, following the semiclassical theory developed for the Bloch system, we construct a wave-packet based on the Floquet-Bloch wavefunctions and thus obtain the effective Lagrangian. The wave-packet is essentially a superposition of wave-forms with a broadened distribution in wavevector, based on which the wave-packet for the electron on a specific Floquet-Bloch band $\omega_\mu$ can be constructed as below:
\begin{equation}\label{wavePacket}
\begin{split}
    &|W_\mu\rangle \equiv \int d\boldsymbol{k} \, a(\boldsymbol{k},\tau_s) \sum_{n,\alpha}f^{\mu,\alpha}_{n,\boldsymbol{k}}(\br_c)|\Phi^\alpha_{n,\boldsymbol{k}}(\br,\br_c,\tau_f)\rangle \\
    &= \sum_{n,\alpha}e^{-in\Omega \tau_f}\int d\boldsymbol{k} \, a(\boldsymbol{k},\tau_s) f^{\mu,\alpha}_{n,\boldsymbol{k}}(\br_c)|\varphi^\alpha_{n,\bk}(\br,\br_c)\rangle
\end{split}
\end{equation}
where $|\varphi^\alpha_{n,\bk}(\br,\br_c)\rangle \equiv e^{i(\boldsymbol{k}+n\boldsymbol{\kappa})\cdot\boldsymbol{r}}|u^\alpha_{n,\bk}(\br,\br_c)\rangle$ is the Bloch wavefunction for the corresponding static system with lattice momentum $\boldsymbol{k}_n \equiv \boldsymbol{k}+n\boldsymbol{\kappa}$ at the Bloch band $E^\alpha$, and $a(\boldsymbol{k},\tau_s)\equiv|a(\boldsymbol{k},\tau_s)|e^{-i\gamma(\boldsymbol{k},\tau_s)}$ is the distribution function in $\boldsymbol{k}$ space. Notice that we put explicitly the packet center $\br_c$ in the wavefunction to illustrate its extra dependence on spatial variation of the external field, whose evaluation is presented in Appendix~\ref{positioncenter}. One of the most important arguments or assumptions we made in the construction of the above wave-packet is that the distribution $a(\boldsymbol{k},\tau_s)$ purely depends on the quasi-momentum $\boldsymbol{k}$ and the slow time scale $\tau_s$. Later we will see that this slow time dependence gives rise to the semiclassical dynamics on a large time scale. The normalization of the constructed wave-packet requires that $\int d\boldsymbol{k}\, |a(\boldsymbol{k},\tau_s)|^2 =1$, and in the sense of wave-packet, we also need the $\boldsymbol{k}$ distribution to be narrow: $|a(\boldsymbol{k},\tau_s)|^2\sim \delta(\boldsymbol{k}-\boldsymbol{k}_c)$ with
\begin{equation}
    \boldsymbol{k}_c (\tau_s) \equiv \int d\boldsymbol{k} \, \boldsymbol{k}|a(\boldsymbol{k},\tau_s)|^2
\end{equation}
being the so-called momentum center of the wave-packet. Again, as we can see, $\boldsymbol{k}_c$ only has slow-time dependence.

Here are some remarks on the construction of the wave-packet using Floquet-Bloch wave functions: (1) Given a Floquet-Bloch system, there will be an infinite number of energy bands in the frequency domain. So, we have to choose one set of Floquet-Bloch bands within the Floquet Brillouin zone, and all others are just replicas labeled by different Floquet index $n$. (2) Similar to the Bloch system, we need adiabaticity for the wave-packet to be well defined, which means that the external slowly varying fields should have energy scales much smaller than the Floquet Gap (which will be defined in the later discussions).

\subsection{Effective Lagrangian and equations of motion}
The effective action for the wave-packet is given by~\cite{sundaram1999wave,kramer1980geometry}
\begin{align}
        \mathcal{S} =& \int dt \langle W_\mu (\bk_c,\br_c)|i\hbar\frac{d}{d t} - \hat{H}(\br,t) | W_\mu(\bk_c,\br_c) \rangle\notag\\
        \approx & \int d\tau_s \langle\langle W_\mu (\bk_c,\br_c)|i\hbar\frac{d}{d t} - \hat{H}(\br,t) | W_\mu(\bk_c,\br_c) \rangle\rangle \notag\\
        \equiv& \int d\tau_s \mathcal{L}_{eff}(\bk_c,\br_c,\tau_s)
\end{align}
where we have implicitly integrated out the fast-time variation and the slow-time dependence $\tau_s$ is left. The fast spatial variation has been taken care of by the inner product between the wave-packets~\cite{sundaram1999wave}. Here, $\mathcal{L}_{eff}(\bk_c,\br_c,\tau_s)$ is defined as the effective Lagrangian,
which has only slow space and time variations. One can also think of the effective Lagrangian as the operator $i\hbar d_t-\hat{H}$ evaluated under the wave-packet over a space-time unit cell.
In Appendix~\ref{time_derivative_wavepacket} and~\ref{energyCorrectionAppendix}, we perform detailed calculations for both $\langle\langle W_\mu|i\hbar d_t |W_\mu\rangle\rangle $ and $\langle\langle W_\mu|\hat{H}|W_\mu\rangle\rangle $.
Thus, we have
\begin{align}\label{Lagrangian}
       &\mathcal{L}_{eff} = -\left(\mathcal{E}_\mu-  \hbar\sum_{n,\alpha}\left| f^{\mu,\alpha}_{n,\boldsymbol{k}_c}\right|^2n\Omega\right)+ \hbar\bk_c\cdot\dot\br_c \notag\\
        & + \left.\langle\langle \Tilde{u}_{\boldsymbol{k}}^{\mu}|i\hbar\partial_{\tau_s}| \Tilde{u}_{\boldsymbol{k}}^{\mu} \rangle\rangle\right|_{\boldsymbol{k}=\boldsymbol{k}_c}+ \hbar\Dot{\boldsymbol{r}}_c\cdot\left.\langle\langle \Tilde{u}_{\boldsymbol{k}}^{\mu}|i\partial_{\boldsymbol{r}_c}| \Tilde{u}_{\boldsymbol{k}}^{\mu} \rangle\rangle\right|_{\boldsymbol{k}=\boldsymbol{k}_c} \notag\\
        &+\hbar\dot\bk_c\cdot\left.\langle\langle \Tilde{u}_{\boldsymbol{k}}^{\mu}|i\partial_{\boldsymbol{k}}| \Tilde{u}_{\boldsymbol{k}}^{\mu} \rangle\rangle\right|_{\boldsymbol{k}=\boldsymbol{k}_c},
\end{align}
where $\mathcal{E}_\mu \equiv \langle\langle W_\mu | \hat{H}(\hat{\boldsymbol{r}},t)| W_\mu \rangle\rangle $ is the energy of the wave-packet and $ | \Tilde{u}_{\boldsymbol{k}}^{\mu}(\br,\br_c) \rangle = \sum_n f^{\mu,\alpha}_{n,\bk}(\br_c)e^{-in\Omega \tau_f}e^{in\boldsymbol{\kappa}\cdot\br}|u^\alpha_{n,\bk}\rangle $ is the periodic part of the Floquet-Bloch wave-function. Note that all fast-time dependencies have been integrated out leaving only slow-time variations. So, here the dot on top of variables $\dot\bO\equiv d\bO/d\tau_s$ means slow time derivative.

Now, by the Euler-Lagrange equation, we can obtain the equations of motion for Floquet-Bloch electrons:
\begin{align}\label{eom}
        \hbar\dot\br_c =& \frac{\partial \tilde{\mathcal{E}}_\mu}{\partial \bk_c} - \left( \vec{\Omega}^\mu_{\bk_c,\bk_c}\cdot\dot\bk_c+ \vec{\Omega}^\mu_{\bk_c,\br_c}\cdot\dot\br_c\right) + \vec{\Omega}^\mu_{\tau_s,\bk_c}; \notag\\
        \hbar\dot\bk_c =& -\frac{\partial \tilde{\mathcal{E}}_\mu}{\partial \br_c} + \left( \vec{\Omega}^\mu_{\br_c,\br_c}\cdot\dot\br_c+ \vec{\Omega}^\mu_{\br_c,\bk_c}\cdot\dot\bk_c\right) - \vec{\Omega}^\mu_{\tau_s,\br_c},
\end{align}
where $ \tilde{\mathcal{E}}_\mu\equiv  \mathcal{E}_\mu-  \hbar\sum_{n,\alpha}\left| f^{\mu,\alpha}_{n,\boldsymbol{k}_c}\right|^2n\Omega = \hbar\omega_\mu + \Delta\mathcal{E}_\mu $ is the electron energy with $\Delta\mathcal{E}_\mu $ being the gradient correction to the energy due to the finite size of the wave-packet (see Appendix~\ref{energyCorrectionAppendix} for more details) and 
\begin{align}
    \left( \vec{\Omega}^\mu_{\bxi,\bxi'} \right)_{p,q} &\equiv \Omega^\mu_{\xi_p,\xi'_q} \equiv i\left[ \langle\langle \partial_{\xi_p}\Tilde{u}_{\boldsymbol{k}}^{\mu}|\partial_{\xi'_q} \Tilde{u}_{\boldsymbol{k}}^{\mu} \rangle\rangle -\text{h.c.}\right], \notag\\
    \left( \vec{\Omega}^\mu_{\tau_s,\bxi} \right)_{p} &\equiv \Omega^\mu_{\tau_s,\xi_p} \equiv i\left[ \langle\langle \partial_{\tau_s}\Tilde{u}_{\boldsymbol{k}}^{\mu}|\partial_{\xi_p} \Tilde{u}_{\boldsymbol{k}}^{\mu} \rangle\rangle -\text{h.c.}\right]
\end{align}
with $\bxi = \bk_c,\br_c$ are the Berry curvatures defined in the parameter space $(\bk_c,\br_c,\tau_s)$.

Now, we have obtained the equation of motion for the Floquet-Bloch electron subjected to slowly varying external potentials. Those equations have exactly the same form as those for the static Bloch crystal~\cite{sundaram1999wave}, except that the quantities involved are modified in the Floquet context, for example, the Berry curvatures that include the time variations in them. In the next few subsections, we are going to discuss some specific physical observables for which one can also find correspondences in the static Bloch crystal together with some implications that are unique to Floquet systems. 

\subsection{Electromagnetic responses of Floquet-Bloch electrons}\label{EM_response}
One of the most important responses studied in condensed matter physics is the electromagnetic response.
In this section, we want to specify the external slow-varying field to be the electromagnetic field, which can be introduced through gauge potentials $[\bA(\br,t),\phi(\br,t)]$:
\begin{equation}
    \hat{H} = \hat{H}_0(\bk+e\bA(\hat{\br},t)) - e\phi(\hat\br,t),
\end{equation}
which gives $\hat{H}_c = \hat{H}_0(\bk+e\bA(\br_c,\tau_s)) - e\phi(\br_c,\tau_s)$. Note that the gauge potentials only introduce slow time and space variations, so that only $(\br_c,\tau_s)$ enters the argument.
Let us use the gauge momentum center $\bq_c = \bk_c+e\bA(\br_c,\tau_s)$, which includes the variation of the position center and the momentum center: $\partial_{\br_c} = (\partial_{\br_c}e\bA)\cdot\partial_{\bq_c}$ and $\partial_{\bk_c} = \partial_{\bq_c}$. Plugging this substitution, we have the electron energy to be
\begin{equation}
    \mathcal{E}_{\mu} = \mathcal{E}_\mu^0(\bq_c) - e\phi(\br_c,\tau_s) + \Delta\mathcal{E}_\mu,
\end{equation}
where $\mathcal{E}_\mu^0(\bq_c)= \langle\langle W_\mu | \hat{H}_0(\bq_c)| W_\mu \rangle\rangle$
and $\Delta\mathcal{E}_\mu =   -\bB\cdot\bm$ (see detailed derivations in Appendix~\ref{energyCorrectionAppendix}),
where $ \bB = \boldsymbol\nabla_{\br_c}\times\bA(\br_c,\tau_s)$ is the magnetic field and 
\begin{equation}\label{magneticmoment}
    \bm = \frac{e}{2\hbar}\Im\langle\langle \frac{\partial}{\partial {\bq_c}}\Tilde{u}_{\bq_c}^{\mu}|\times\left[\hat{H}^F(\bq_c)-\hbar\omega_\mu(\bq_c)\right]|\frac{\partial}{\partial\bq_c}\Tilde{u}_{\bq_c}^{\mu}\rangle\rangle
\end{equation}
is the local magnetic moment of the Floquet-Bloch wave-packet due to its self-rotation in the magnetic field.
Finally, the Lagrangian becomes
\begin{align}
        \mathcal{L}^{EM}_{eff} 
        = & -\mathcal{E}_\mu^M + e\phi(\br_c,\tau_s) + \hbar\dot\br_c\cdot[\bq_c-e\bA(\br_c,\tau_s)] \notag\\
        &+ \hbar\dot\bq_c\cdot\langle\langle \Tilde{u}_{\bq_c}^{\mu}|i\partial_{\bq_c}| \Tilde{u}_{\bq_c}^{\mu} \rangle\rangle
\end{align}
where $\mathcal{E}_\mu^M\equiv \hbar\omega_\mu(\bq_c)-\bm\cdot\bB$ is the electron energy including the magnetization, and the last term is equal to the last three terms in Eq.~\eqref{Lagrangian} writing compactly. The equations of motion become simply
\begin{equation}
    \dot\br_c = \frac{\partial \mathcal{E}_\mu^M}{\hbar\bq_c} - \dot\bq_c\times\vec{\Omega}; \qquad \dot\bq_c = -\frac{e\vec E}{\hbar} - e\dot\br_c\times\bB
\end{equation}
where $(\vec\Omega)_p = \frac{1}{2}\epsilon_{ijk}\Omega_{q_{cj},q_{ck}}$ is pseudo vector form of the Berry curvature and $\vec E = -\vec\nabla_{\br_c}\phi(\br_c,\tau_s)-\partial_{\tau_s}\bA(\br_c,\tau_s)$ is the electric field.

The electromagnetic fields introduced here should be slowly varying in both space and time. This is important since, in some cases, the Floquet system is actually driven by electromagnetic fields which introduce the fast time variations~\cite{morimoto2016topological,gao2022dc}. One has to distinguish those fields by their time scales. The general argument here is still the separation of time scales.

\subsection{Berry curvature contribution to the magnetization}
Apart from the local magnetic moment of the wave-packet as written in Eq.~\eqref{magneticmoment}, there is another contribution to the overall magnetization originating from the Berry curvature. The Bloch counterpart can be found in Ref.~\cite{xiao2010berry}. Similarly, it can be seen from the current response of the Floquet-Bloch system:
\begin{align}\label{current_response_general}
	\boldsymbol j_\mu =& -e\int_{BZ} d\bk \rho^F_{\mu,\mu}\dot{\br_c} = -\frac{e}{\hbar}\int_{BZ} d\bk 
	\Big[\frac{\partial \mathcal{E}_\mu}{\partial \bk_c}(1+\text{Tr }\boldsymbol\Omega^\mu_{\bk \br}) \notag\\
 &-\boldsymbol\Omega^\mu_{\bk \br}\cdot\dot{\br_c}-\boldsymbol\Omega^\mu_{\bk \bk }\cdot\dot{\bk_c}+\boldsymbol\Omega^\mu_{\tau_s\bk}\Big],
\end{align}
where for simplicity we set $\rho^F_{\mu,\mu}\equiv 1$, i.e., a fully occupied band, and $\text{Tr }\boldsymbol\Omega^\mu_{\bk \br}$ is a correction to the $\bk$-space measure $d\bk$ (or the density of state) due to the non-canonical form of the equations of motions~\cite{xiao2005berry,xiao2010berry}. In this case, we also consider no external slow time variations: $\boldsymbol\Omega^\mu_{\tau_s\bk}=0 $, and $\dot{\bk_c}=-\frac{\partial \mathcal{E}_\mu}{\partial \br_c}$, $\dot{\br_c}=\frac{\partial \mathcal{E}_\mu}{\partial \bk_c}$ (notice that any higher orders in the Berry curvatures have been ignored). Thus, the current can be rewritten as (by discarding terms that are total derivatives of momentum $\bk$)
\begin{align}
        &\boldsymbol j_\mu = -\frac{e}{\hbar}\int_{BZ} d\bk 
	\mathcal{E}_\mu\Big[-\frac{\partial }{\partial \bk}\boldsymbol\Omega^\mu_{k_i r_i} + \frac{\partial }{\partial k_i}\boldsymbol\Omega^\mu_{\bk r_i} - \frac{\partial }{\partial r_i}\boldsymbol\Omega^\mu_{\bk k_i}\Big] \notag\\
	&-\frac{e}{\hbar}\frac{\partial }{\partial r_i}\int_{BZ} d\bk 
	\mathcal{E}_\mu\boldsymbol\Omega^\mu_{\bk k_i} = \boldsymbol{\nabla}\times \int_{BZ} d\bk 
	(-\frac{e}{\hbar})\mathcal{E}_\mu\boldsymbol\Omega^\mu,
\end{align}
where the repeated subscript ``$i$" should be summed over (no summation for index $\mu$) and we have recognized that the first line is exactly the Bianchi identity which is zero (one can also show explicitly that those terms cancel each other exactly). Here $\boldsymbol\Omega^\mu$ is again the pseudo vector form of the Berry curvature. Finally, the current can be cast into a curl of something that has the meaning of magnetization. Therefore, we conclude that there is a contribution from the Berry curvature to the total magnetization which reads
\begin{equation}
    -\frac{e}{\hbar}\mathcal{E}_\mu\boldsymbol\Omega^\mu = e\omega_\mu(\bk)\Im\langle\langle \frac{\partial}{\partial {\bk}}\Tilde{u}_{\bk}^{\mu}|\times|\frac{\partial}{\partial\bk}\Tilde{u}_{\bk}^{\mu}\rangle\rangle.
\end{equation}
Together with the local magnetic moment~\eqref{magneticmoment}, the Magnetization at given $\bk$-state becomes (promoting $\bq_c$ to general $\bk$ here)
\begin{equation}\label{total_magnetization}
    \bM(\bk) = \frac{e}{2\hbar}\Im\langle\langle \frac{\partial}{\partial {\bk}}\Tilde{u}_{\bk}^{\mu}|\times\left[\hat{H}^F(\bk)+\hbar\omega_\mu(\bk)\right]|\frac{\partial}{\partial\bk}\Tilde{u}_{\bk}^{\mu}\rangle\rangle.
\end{equation}

Some remarks need to be given regarding the above derivations. First, the procedure for getting the Berry curvature contribution to the total magnetization is general for any systems that have the current response as described in Eq.~\eqref{current_response_general}. So, the obtained magnetization in Eq.~\eqref{total_magnetization} is also true for static Bloch systems at zero temperature (because we set $\rho^F_{\mu,\mu}\equiv 1$)~\cite{xiao2005berry,shi2007quantum} but modifications in energy have to be made: $H\to H-\mu$ and $\hbar\omega \to \hbar\omega -\mu$ to account for the chemical potential~\cite{Note2}. In this section, we set simply the distribution function to be constant which ignores the effect of $\bk$-dependence of the Floquet-Bloch density of states (as we will discuss later). Such contribution from the non-uniform distribution function can be found quite easily in Bloch systems~\cite{xiao2010berry} but becomes complicated in the Floquet system due to its non-equilibrium nature, which is beyond our scope in this work. However, Efforts have been given to address that issue using perturbation theory~\cite{topp2022orbital}.

\subsection{Effective Hamiltonian and the averaged energy}
Here we want to provide some insights into the semiclassical dynamics derived above. Given a system with two time scales: fast periodic time evolution and external slow time evolution, we separate those two by imposing independence between them and average out the fast-time ending up with only the dynamics of slowly varying quantities $(\boldsymbol{r}_c,\boldsymbol{k}_c,\tau_s; \boldsymbol{E}, \boldsymbol{B},\cdots)$. On the large time scale, we can define the effective Hamiltonian based on only those slowly varying variables. Firstly, from Lagrangian Eq.\eqref{Lagrangian}, we have the canonical conjugate variables:
\begin{align}
        \boldsymbol{\Pi_{\boldsymbol{r}_c}} &= \frac{\partial \mathcal{L}_{eff}}{\partial \Dot{\boldsymbol{r}}_c} = \hbar\boldsymbol{k}_c +  \hbar\langle\langle \Tilde{u}_{\boldsymbol{k}_c}^{\mu}|i\partial_{\boldsymbol{r}_c}| \Tilde{u}_{\boldsymbol{k}_c}^{\mu} \rangle\rangle \notag\\
        \boldsymbol{\Pi_{\boldsymbol{k}_c}} &= \frac{\partial \mathcal{L}_{eff}}{\partial \Dot{\boldsymbol{k}}_c} = \hbar\langle\langle \Tilde{u}_{\boldsymbol{k}_c}^{\mu}|i\partial_{\boldsymbol{k}_c}| \Tilde{u}_{\boldsymbol{k}_c}^{\mu} \rangle\rangle.
\end{align}
Then, the effective Hamiltonian is obtained through Legendre transformation:
\begin{align}
    &\mathcal{H}_{eff}(\boldsymbol{r}_c,\boldsymbol{k}_c,\boldsymbol{\Pi_{\boldsymbol{r}_c}},\boldsymbol{\Pi_{\boldsymbol{k}_c}},\tau_s) = \Dot{\boldsymbol{r}}_c\cdot\boldsymbol{\Pi_{\boldsymbol{r}_c}} + \Dot{\boldsymbol{k}}_c\cdot\boldsymbol{\Pi_{\boldsymbol{k}_c}} -\mathcal{L}_{eff} \notag\\
    &= \mathcal{E}_\mu -\hbar\sum_{n,\alpha}\left| f^{\mu,\alpha}_{n,\boldsymbol{k}_c} \right|^2 n\Omega -\hbar\langle\langle \Tilde{u}_{\boldsymbol{k}_c}^{\mu}|i\partial_{\tau_s}| \Tilde{u}_{\boldsymbol{k}_c}^{\mu} \rangle\rangle.
\end{align}
As we can see, the effective Hamiltonian is generically time($\tau_s$)-dependent due to the last term induced by the external slow time-dependent field. Let us consider the simplest case where there is no external field, the Hamiltonian becomes $\mathcal{H}_{eff} = \langle\langle H^F_0\rangle\rangle -\hbar\sum_{n,\alpha}\left| f^{\mu,\alpha}_{n,\boldsymbol{k}_c} \right|^2 n\Omega$ that is invariant under time ($\tau_s$) translation: $\frac{d}{d\tau_s}\mathcal{H}_{eff} = \frac{\partial}{\partial\tau_s}\mathcal{H}_{eff} = 0$. In the case where there is no external field, we have $\Dot{\boldsymbol{k}}_c = 0$ and $\frac{\partial}{\partial\boldsymbol{r}_c}f^{\mu,\alpha}_{n,\boldsymbol{k}_c} = 0$, which leads to $\frac{d}{d\tau_s} \left| f^{\mu,\alpha}_{n,\boldsymbol{k}_c} \right|^2= (\Dot{\boldsymbol{r}}_c\cdot\frac{\partial}{\partial\boldsymbol{r}_c} + \Dot{\boldsymbol{k}}_c\cdot\frac{\partial}{\partial\boldsymbol{k}_c})\left| f^{\mu,\alpha}_{n,\boldsymbol{k}_c} \right|^2 = 0$. Together, we have the conserved Energy:
\begin{equation}
    \frac{d}{d\tau_s}\langle\langle H^F_0\rangle\rangle = 0
\end{equation}
in the sense of expectation energy averaged over a Floquet period. Here $H^F_0$ stands for the Floquet-Bloch system without an external field. Its expectation value has been evaluated in Eq.~\eqref{leadingOrderEnergy}, which contains two terms: the electron quasi-energy and the additional energy related to the Floquet drive with frequency $\Omega$. The latter can be viewed as the driving field quanta ($n\hbar\Omega$) coherently coupled to the electron. For example, if the system is driven by coherent light, then the additional energy will be the energy of the photon coupled to the electron. The average number of the photon coupled to the electron at state $\omega_\mu(\bk)$ is $\sum_{n,\alpha}\left| f^{\mu,\alpha}_{n,\boldsymbol{k}} \right|^2 n$.

More intriguingly, if we consider an external field with only space variation such as electric field: $H = H_0(\boldsymbol{k},t) - e\phi(\boldsymbol{r})$, then the effective Hamiltonian is still conserved but now becomes 
\begin{equation}
    \mathcal{H}_{eff} = \langle\langle H^F_0\rangle\rangle -e\phi(\boldsymbol{r}_c) - \hbar\sum_{n,\alpha}\left| f^{\mu,\alpha}_{n,\boldsymbol{k}_c} \right|^2 n\Omega.
\end{equation}
Therefore, we have the energy conversion between three sources:
\begin{equation}\label{energyConversion}
    0 = \frac{d}{d\tau_s}\mathcal{H}_{eff} \equiv \frac{d}{d\tau_s}\langle\langle H^F_0\rangle\rangle + \frac{d}{d\tau_s}E_{\text{pot}} + \frac{d}{d\tau_s} N(\tau_s)\hbar\Omega,
\end{equation}
where $\frac{d}{d\tau_s}N(\tau_s) \equiv -\frac{d}{d\tau_s}\sum_{n,\alpha}n\left| f^{\mu,\alpha}_{n,\boldsymbol{k}_c} \right|^2$. Thus, we can identify that
\begin{enumerate}
    \item $\langle\langle H^F_0\rangle\rangle$ is the Floquet-Bloch energy of electron defined as the averaged energy without external potential. It contains the energy of the electron and the energy of Floquet coherent driven field (such as the photon, $\hbar\Omega$);
    \item $E_{\text{pot}}\equiv -e\phi(\boldsymbol{r}_c)$ is the potential energy of the electron subjected to the external field;
    \item $N(\tau_s)\hbar\Omega$ is the energy of the driven source of the Floquet system.
\end{enumerate}
We have recognized the last changing term $N(\tau_s)\hbar\Omega$ as the driven source term because as defined in Eq.~\eqref{energyConversion}, $-\frac{d}{d\tau_s}N(\tau_s) = \frac{d}{d\tau_s}\sum_{n,\alpha}n\left| f^{\mu,\alpha}_{n,\boldsymbol{k}_c} \right|^2 $, the $N(\tau_s)$ is to exactly compensate the change in the number of driving field quanta that are coherently coupled to the electron.  

We want to remark here that the effective Hamiltonian gives us a hint to construct an energetic conserved quantity for Floquet systems and allows us to discuss the role of the Floquet driving field (not to be confused with the external field) in energy conversion processes.

\subsection{Polarization of a fully occupied Floquet-Bloch band}
In this section, we want to have a general discussion on the polarization of a space-time crystal.
To have a well-defined polarization, we have to consider a space-time insulator or a fully occupied Floquet-Bloch band with electron distribution: $\rho^{F}_{\mu,\mu}(\bk,t)=1$ for all $\bk$. For the equations of motion given by Eq.~\eqref{eom}, we can find the polarization of the space-time crystal through adiabatic current~\cite{Note3},
i.e., letting $\boldsymbol\Omega^\mu_{\bk \br}=0$ and $\dot{\bk_c}=0$ in Eq.~\eqref{current_response_general}, so that the current reduces to (from now on, we omit the electric charge $-e$ and set $\hbar=1$ for simplicity)
\begin{equation}
	\boldsymbol j_\mu = \int_{BZ} d\bk 
	\left(\frac{\partial \mathcal{E}_\mu}{\partial \bk_c}+\boldsymbol\Omega^\mu_{\tau_s \bk}\right) 
	=\int_{BZ} d\bk  \boldsymbol\Omega^\mu_{\tau_s \bk} 
\end{equation}
where we utilized the periodicity of $\mathcal{E}_\mu$. Therefore the polarization due to the current accumulation is 
\begin{align}
	&\Delta \boldsymbol P_\mu = \int d\tau_s \,\boldsymbol j_\mu = \int d\tau_s \int_{BZ} d\bk  \boldsymbol\Omega^\mu_{\tau_s \bk} \notag\\
 &= \int d\tau_s \int_{BZ} d\bk \left[\partial_{\tau_s}\tilde{\mathcal{A}}^\mu_{\bk}-\partial_{\bk_c}\tilde{\mathcal{A}}^\mu_{\tau_s}\right]= \left[ \int_{BZ} d\bk\tilde{\mathcal{A}}^\mu_{\bk} \right]_{\tau_{s_1}}^{\tau_{s_2}},
\end{align}
where $ \tilde{\mathcal{A}}^\mu_{\bk(\tau_s)} = \langle\langle \tilde{u}^\mu_\bk |i\partial_{\bk(\tau_s)}|\tilde{u}^\mu_\bk\rangle\rangle$ and we have chosen a periodic gauge for $\tilde{\mathcal{A}}^\mu_{\tau_s}$.
Accordingly, we can define the adiabatic polarization as
\begin{equation}
	\boldsymbol P_\mu = \int_{BZ} d\bk\tilde{\mathcal{A}}^\mu_{\bk}.
\end{equation}
Explicitly, we have
\begin{align}
	&\tilde{\mathcal{A}}^\mu_{\bk} = \langle\langle \tilde{u}^\mu_\bk |i\partial_{\bk}|\tilde{u}^\mu_\bk\rangle\rangle \notag\\
 &\equiv \sum_{n}\left[ \sum_{\alpha,\beta}\left( f^{\mu,\alpha}_{n,\bk} \right)^*f^{\mu,\beta}_{n,\bk} \mathcal{A}^{\alpha\beta}_{\bk+n\boldsymbol\kappa} + \sum_{\alpha} i \left( f^{\mu,\alpha}_{n,\bk} \right)^*\partial_{\bk}f^{\mu,\alpha}_{n,\bk}\right],
\end{align}
where $ \mathcal{A}^{\alpha\beta}_{\bk}$ is the interband Berry connection for the original Bloch crystal.
Thus, the polarization will have two parts:
\begin{align}
	\boldsymbol P &= \sum_{n}\int_{BZ} d\bk\left[ \sum_{\alpha,\beta}\left( f^{\mu,\alpha}_{n,\bk} \right)^*f^{\mu,\beta}_{n,\bk} \mathcal{A}^{\alpha\beta}_{\bk+n\boldsymbol\kappa}\right. \notag\\
 &\quad+ \left.\sum_{\alpha} i \left( f^{\mu,\alpha}_{n,\bk} \right)^*\partial_{\bk}f^{\mu,\alpha}_{n,\bk}\right]\equiv \sum_{n}\boldsymbol{P}_n
\end{align}
where the first term mainly originates from Bloch contribution but is modified by the time perturbation (weighted by the factor $\left( f^{\mu,\alpha}_{n,\bk} \right)^*f^{\mu,\beta}_{n,\bk}$), while the second term is the purely time-periodic effect. Let's call the first term the Bloch contribution and the second term the Floquet contribution. Here, we denote $\boldsymbol{P}_n$ as the partial polarization at the $n$-th Floquet sub-band (or replica).

It is important that the physical quantities or observables should be gauge-invariant. Consider the following gauge transformation: $|\tilde{u}^\mu_\bk\rangle\mapsto e^{-i\theta^\mu_\bk}|\tilde{u}^\mu_\bk\rangle $ or equivalently $f^{\mu,\alpha}_{n,\bk}\mapsto e^{-i\theta^\mu_\bk}f^{\mu,\alpha}_{n,\bk}$, the polarization $\boldsymbol{P}$ will gain an extra term $ \int_{BZ} d\bk \partial_{\bk}\theta^\mu_\bk$ which vanishes if we impose periodic boundary conditions for the wave function. However, the partial polarization will become gauge-dependent, thus it cannot be a measurable quantity.  

Before we close this section, we would like to discuss special symmetry constrain that will force the Floquet contribution to be zero. For oblique space-time crystals with $\boldsymbol{\kappa}\neq 0$, we have that the parity and time reversal symmetry are both broken by the Floquet driven which selects unique directions in both space and time. However, the system can still preserve the joint spatial and temporal reversal symmetry called the $\mathcal{PT}$ symmetry, which requires that the Hamiltonian is invariant under the joint action of reversing both space and time. Typically, we should have Hamiltonian in the following form:
\begin{equation}
    \hat{H}^F(\br,t) = -\frac{\hbar^2}{2m}\partial^2_{\br} + V_B(\br) + V_F(\br,t) - i\hbar\partial_{t}
\end{equation}
where $V_B(\br)$ is the periodic potential of the original Bloch crystal and $V_F(\br,t)$ is the Floquet drive. Those two potentials are all real-valued.
Under the $\mathcal{PT}$ action, we have
\begin{equation}
    \mathcal{PT}\hat{H}^F(\br,t)(\mathcal{PT})^{-1} = \hat{H}^{F*}(-\br,-t) = \hat{H}^F(\br,t)
\end{equation}
which requires that $V_B(\br) = V_B(-\br)$ (this implies that the periodic part of the Bloch wavefunction can be chosen to be: $|u^\alpha_{\boldsymbol{k}}(\boldsymbol{r})\rangle = |u^\alpha_{\boldsymbol{k}}(-\boldsymbol{r})\rangle$) and $V_F(\br,t) = V_F(-\br,-t) $. Notice that $\mathcal{PT}\bk(\mathcal{PT})^{-1} = \bk$, thus we have the extended Floquet-Bloch basis in Eq.~\eqref{FBeigenstate} is also $\mathcal{PT}$ symmetric: 
\begin{equation}
    \mathcal{PT}|\Phi^\alpha_{n,\boldsymbol{k}}(\boldsymbol{r},t)\rangle =|\Phi^{\alpha }_{n,\boldsymbol{k}}(-\boldsymbol{r},-t)\rangle^*=|\Phi^\alpha_{n,\boldsymbol{k}}(\boldsymbol{r},t)\rangle.
\end{equation}
Then, for the Floquet-Bloch eigenstate to be $\mathcal{PT}$ symmetric, we require that the coefficient $f^{\mu,\alpha}_{n,\bk}$ to be real for all $n,\alpha$ up to an overall gauge choice. One should see that the contribution in the polarization from the Floquet variation vanishes identically due to the fact that $ \sum_{n,\alpha}|f^{\mu,\alpha}_{n,\bk}|^2 = 1$. However, the modified Bloch contribution to the polarization may still survive under the $\mathcal{PT}$ symmetry.

\section{Floquet quasi-equilibrium}\label{densityMatrixAnalysisSection}
In the previous section, we developed the single electron dynamics in Floquet-Bloch systems which show similar behavior as in Bloch systems but are modified by Floquet time variation. The next step is naturally to consider an ensemble of electrons (but still non-interacting for simpleness). However, such consideration turns out to be difficult, given that the Floquet-Bloch systems are essentially out-of-equilibrium. There have been many efforts dedicated to exploring the non-equilibrium behavior of electrons in Floquet systems~\cite{dehghani2015out,seetharam2015controlled,iadecola2015floquet,Shirai2015Condition,sato2020floquet}. However, we can still bypass the difficulty of dealing with non-equilibrium ensembles by considering a special Floquet quasi-equilibrium ensemble~\cite{gao2022dc}, where the density matrix becomes time-independent and diagonal under the Floquet-Bloch eigenbasis just resembling a Bloch equilibrium ensemble. In this section, we follow the formalism introduced in Ref.~\cite{gao2022dc} but extend it to the case of finite $\boldsymbol{\kappa}$, i.e., the oblique space-time crystal.

\subsection{General formalism}
To know the physical quantities contributed by many (non-interacting) electrons, we need to know the population or density of electrons at different states. We start with the Liouville equation 
\begin{equation}\label{LiouvilleEq}
    i\partial_{t} \hat{\rho} = [\hat{H},\hat{\rho}] + i[\hat{D},\hat{\rho}],
\end{equation}
where $\hat{\rho}$ is the density operator of the Floquet-Bloch system and $\hat{D}$ characterizes the damping. Under the relaxation time approximation and assuming non-interacting electrons, we can write the damping term as~\cite{gao2022dc}
\begin{equation}
    [\hat{D},\hat{\rho}] = -\Gamma \left( \hat{\rho} - \hat{\rho}^{B,eq} \right),
\end{equation}
where $\Gamma = 1/\tau_{s_r}$ is the damping constant with $\tau_{s_r}$ being the relaxation time and $ \hat{\rho}^{B,eq}$ is the equilibrium density operator of the corresponding static Bloch system. Although the damping term introduced above is oversimplified so that some interacting or correlation features of electrons are ignored, we believe that such a simple setting can already give the physics that we want to address in this work, and further treatment is certainly needed when including electron-electron interactions.
For the rest of the discussion, we assume that the density operator of the Floquet-Bloch system is quasi-momentum-diagonal if the system is in a steady state. However, the oblique space-time crystal ($\boldsymbol{\kappa}\neq 0$) is essentially different from the rectangular space-time crystal ($\boldsymbol{\kappa}=0$), given that the quasi-momentum in oblique space-time crystals is only conserved up to an integer multiple of $\boldsymbol{\kappa}$, i.e., the electron can acquire momenta from the Floquet drive. In this case, the best we can do is to define the density operator in the following way:
\begin{align}\label{DensityOperator_OSTC}
    \hat{\rho} =& \sum_{\mu,n,\nu,m}\int' d\bk \rho^{F}_{\{\mu,n\},\{\nu,m\}}(\bk,t)\notag\\
    &\qquad\times|\Psi^{\mu}_{\bk+n\boldsymbol{\kappa}}(\br,t)\rangle\langle\Psi^\nu_{\bk+m\boldsymbol{\kappa}}(\br,t)|
\end{align}
where $ \rho^{F}_{\{\mu,n\},\{\nu,m\}}(\bk,t)\equiv \langle\Psi^\mu_{\bk+n\boldsymbol{\kappa}}(\br,t)|\hat{\rho}|\Psi^\nu_{\bk+m\boldsymbol{\kappa}}(\br,t)\rangle$. One has to be very careful about the summation over indices $n,m$ and the integration over the quasi-momentum $\bk$ to avoid repeated counting for the same states. A detailed discussion regarding the integration and summation can be found in Appendix~\ref{change_integration}. For the static Bloch equilibrium density operator, we want to keep it in general which reads
\begin{equation}
    \hat{\rho}^{B,eq} = \sum_{\alpha,\beta} \int d\bk d\bk' \rho^{B,eq}_{\alpha,\beta}(\bk,\bk')|\varphi^\alpha_\bk(\br)\rangle \langle \varphi^\beta_{\bk'}(\br)|,
\end{equation}
where $\rho^{B,eq}_{\alpha,\beta}(\bk,\bk') \equiv  \langle \varphi^\alpha_{\bk}(\br)|\hat{\rho}^{B,eq}|\varphi^\beta_{\bk'}(\br)\rangle$ is the density matrix of the static Bloch equilibrium under the Bloch eigenbasis.

By casting Eq.~\eqref{LiouvilleEq} into the Floquet-Bloch basis, after some algebra (see Appendix~\ref{densityMatrixFordamping}),
we can obtain:
\begin{align}\label{densityMatrix}
    &\rho^{F,l}_{\{\mu,n\},\{\nu,m\}}(\bk)=i\Gamma\delta_{n-m,l}\times \notag\\
    & \sum_{\alpha\beta,p}\frac{\rho^{B,eq}_{\alpha,\beta}(\bk+n\boldsymbol{\kappa}+p\boldsymbol\kappa)\left( f^{\mu,\alpha}_{p,\bk+n\boldsymbol{\kappa}} \right)^*f^{\nu,\beta}_{p+l,\bk+m\boldsymbol{\kappa}}}{\omega^n_\mu(\bk)-\omega^m_\nu(\bk)-i\Gamma}.
\end{align}
Here $\omega_\mu^n(\bk)\equiv\omega_\mu(\bk+l\boldsymbol{\kappa})-n\Omega$ corresponds to the $\mu$-th Floquet-Bloch band at the $n$-th Floquet replica. One can think about this as an extended zero picture in the frequency domain~\cite{gao2021floquet}.
Notice that we have $\lim_{\Gamma\to 0}(-i\Gamma)/(\omega^n_\mu(\bk)-\omega^m_\nu(\bk)-i\Gamma) = \delta_{\mu,\nu}\delta_{n,m}$ if the system is gapped, then when the thermal contact is gradually turned off ($\Gamma \to 0$), the density matrix becomes
\begin{equation}\label{diagonalDensityMatrix}
\begin{split}
    &\rho^{F}_{\{\mu,n\},\{\nu,m\}}(\bk,t) = \delta_{\mu,\nu}\delta_{n,m}\\
    &\times\sum_{\alpha\beta,p}\rho^{B,eq}_{\alpha,\beta}(\bk+n\boldsymbol\kappa+p\boldsymbol{\kappa})\left( f^{\mu,\alpha}_{p,\bk+n\boldsymbol\kappa} \right)^*f^{\mu,\beta}_{p,\bk+n\boldsymbol\kappa},
\end{split}
\end{equation}
which is diagonal and time independent. Thus, we would like to call it the Floquet quasi-equilibrium state to emphasize the diagonal and time-independent density matrix under the Floquet-Bloch eigenbasis. The density operator can then be written as
\begin{align}
        \hat{\rho} &= \sum_{\mu,n}\int' d\bk \rho^{F}_{\{\mu,n\},\{\mu,n\}}(\bk,t)|\Psi^{\mu}_{\bk+n\boldsymbol{\kappa}}(\br,t)\rangle\langle\Psi^\mu_{\bk+n\boldsymbol{\kappa}}(\br,t)| \notag\\
        &= \sum_{\mu}\int_{BZ} d\bk \rho^{F}_{\mu,\mu}(\bk)|\Psi^{\mu}_{\bk}(\br,t)\rangle\langle\Psi^\mu_{\bk}(\br,t)|,
\end{align}
where the change of the integration has been discussed in Appendix~\ref{change_integration} and
\begin{align}\label{diagonal_densityMatrix}
    \rho^{F}_{\mu,\mu}(\bk) &= \rho^{F}_{\{\mu,n=0\},\{\mu,n=0\}}(\bk) \notag\\
    &= \sum_{\alpha\beta,p}\rho^{B,eq}_{\alpha,\beta}(\bk+p\boldsymbol{\kappa})\left( f^{\mu,\alpha}_{p,\bk} \right)^*f^{\mu,\beta}_{p,\bk}.
\end{align}
From the above density matrix element, we can find that it is not easy to form a Floquet-Bloch insulator. One of the possible ways to do that is to let the original Bloch bands involved in forming the Floquet-Bloch bands all be occupied, namely $\rho^{B,eq}_{\alpha,\beta}(\bk) =\delta_{\alpha,\beta}$, thus, we can have $\rho^{F}_{\mu,\mu}(\bk)=1$.

Now, given an observable $\hat{O}$, we can say that the ensemble average of the expectation value is simply
\begin{equation}\label{expectationValue}
    \langle\langle \hat{O} \rangle\rangle = \Tr^{F}[\hat{\rho}\hat{O}] = \sum_{\mu}\rho^{F}_{\mu,\mu}\langle\langle\Psi^\mu_\bk(\br,t)|\hat{O}|\Psi^\mu_\bk(\br,t)\rangle\rangle,
\end{equation}
where $\Tr^{F}[\hat{O}] \equiv \sum_{\mu}\langle\langle\Psi^\mu_\bk(\br,t)|\hat{O}|\Psi^\mu_\bk(\br,t)\rangle\rangle$ is the space-time trace under the Floquet-Bloch eigenbasis. We want to emphasize here that we only consider the intrinsic contribution of the density matrix, i.e. at the condition where $\Gamma \to 0$. For finite $\Gamma$, we shall expect off-diagonal (or extrinsic) contributions. We want to emphasize here that Eq.~\eqref{expectationValue} is the key to why the semiclassical analysis works for Floquet-Bloch systems.

Finally, we would like to point out that the analysis in this section represents a special initialization protocol (connecting the Floquet-Bloch system to a heat bath and then gradually turning off the connection), following which the Floquet-Bloch system can be prepared into such a state with the density matrix in Eq.~\eqref{diagonal_densityMatrix}.

\subsection{Finite damping: a discussion}
There rises a natural question when damping is finite: does the semiclassical scheme developed for intrinsic Floquet-Bloch systems still work? To answer this question, one should be clear about what is needed to validate the semiclassical theory. An observation is that the semiclassical formalism discussed in Sec.~\ref{semiclassical_dynamics} only involves single-electron states while the many-electron effects are restored by simply multiplying the single-electron dynamics by their densities or populations. This then requires that the density operator of the system (without any scattering or interactions) should be diagonal in the basis on which the wave-packet is constructed. Moreover, for Floquet-Bloch systems, the density matrix also needs to be fast-time-independent as we have already averaged over fast-time variations of the single-electron dynamics~\cite{Note4}. Having the above criteria in mind, we will show that it is impossible to apply our semiclassical theory to damping Floquet-Bloch systems and that further development is needed.

First of all, it is easy to see why the semiclassical formalism fails when $\boldsymbol{\kappa}\neq 0$. In Sec.~\ref{densityMatrixAnalysisSection}, we have shown that the off-diagonal term in the density matrix for non-zero damping involves different states with different quasi-momenta $\bk+n\boldsymbol{\kappa}$ and $\bk+m\boldsymbol{\kappa}$ ($n\neq m$) which violates the assumption used for constructing the wave-packet in Eq.~\eqref{wavePacket} where only one specific quasi-momentum is involved. This is a unique reason for the oblique space-time crystal which does not hold for the rectangular space-time crystal where $\kappa=0$. However, for $\kappa=0$, we can show that the density matrix can be written in a diagonal form (see Appendix~\ref{densitymatrixrectangular}):
\begin{align}
    &\rho^F_{\mu,\nu}(\bk,t;\Gamma) = \delta_{\mu,\nu}\times\notag\\
    &\left[ \rho^F_{\mu,\mu}(\bk) + i\Gamma\sum_{\alpha\beta,n}\sum_{l\neq 0}\frac{\rho^{B,eq}_{\alpha,\beta}(\bk)\left( f^{\mu,\alpha}_{n,\bk} \right)^*f^{\mu,\beta}_{n+l,\bk}}{l\Omega+i\Gamma}e^{-il\Omega t} \right],
\end{align}
which will always contain fast-time-dependent parts. In general, there is no such orthonormal basis that can make the density matrix diagonal and time-independent simultaneously for a non-zero $\Gamma$. As we discussed at the beginning of this section, for density matrices having fast-time variations, the semiclassical analysis fails, given that Eq.~\eqref{expectationValue} does not hold anymore.

Before the end of this whole Sec.~\ref{densityMatrixAnalysisSection}, we want to make the following comment. Although the damping profile introduced to study the density matrix is oversimplified as a single relaxation time approximation where many other correlated effects have been ignored, we believe that the arguments stated in this section work for general damped Floquet-Bloch systems, especially when the damping term is sufficiently small. The key ingredient of our analyses (both the semiclassical theory and the density matrix we investigated here) is the assumption of non-interacting electrons which may not be true in strongly correlated systems. Special cares have to be taken to incorporate the electron interactions or correlations into the theory.

So far, we have derived single electron behavior and also the electronic density in a Floquet-Bloch band, to show how those two combined together in studying the response of such a system, in the next section, we will study a simple but physically meaningful example, showcasing its unique current responses.

\section{Intrinsic DC Current Response in Oblique Space-time Crystal}\label{caseStudy}
After obtaining the density matrix of the Floquet quasi-equilibrium, we are now able to discuss many responses in such equilibrium:
\begin{equation}
    \chi[\hat{O}] = \sum_\mu\int d\bk\rho^F_{\mu,\mu}(\bk)  \langle\langle\Psi^\mu_\bk(\br,t)|\hat{O}|\Psi^\mu_\bk(\br,t)\rangle\rangle,
\end{equation}
where the operator $\hat{O}$ can have various possibilities. For example, we can calculate the charge current for $\hat{O}=-e\hat{\bv}$, the spin current for $\hat{O}=\sigma\hat{\bv}$, and/or the magnetization for $\hat{O}=-e\hat{\br}\times\hat{\bv}$. The semiclassical theory developed in Sec.~\ref{semiclassical_dynamics} provides a systematic way of evaluating the expectation value $\langle\langle\hat{O}\rangle\rangle$ even in the presence of external slowly-varying fields. In this section, we consider the intrinsic charge current response in oblique space-time crystals as a simple application, while more complicated cases are left to future works. 

Recall the general form of the current response in Eq.~\eqref{current_response_general}, without any external field, we can simplify it to be
\begin{equation}\label{intrinsicCurrent}
\begin{split}
    \boldsymbol{J}^{\text{In}} &= -e\sum_{\mu}\int d\bk\rho^F_{\mu,\mu}(\bk)\dot{\br}_c = -e\sum_{\mu}\int d\bk \rho^F_{\mu,\mu}(\bk)\frac{\partial\omega_\mu(\bk)}{\partial\bk}.
\end{split}
\end{equation}
This expression, although very simple, is highly non-trivial in an oblique space-time crystal. In an ordinary Bloch system reaching a thermal equilibrium state, the intrinsic current must vanish for two reasons: the distribution function must be a function of the energy and the energy is a periodic function of momentum $\bk$, which do not necessarily hold true for Floquet-Bloch systems given that they are essentially non-equilibrium. However, we will need more conditions for getting non-zero intrinsic current in a Floquet-Bloch system, and this is why we consider an oblique space-time crystal.

In general, for this current to be non-zero, we require the following: (1) the inversion symmetry and the time-reversal symmetry have to be broken otherwise the contributions from $\bk$ and $-\bk$ will cancel each other exactly; (2) more importantly, the electron population $\rho^F_{\mu,\mu}(\bk)$ can not be an analytical function of the energy $\omega_\mu(\bk)$, otherwise, the integration will always give zero due to the periodicity of the spectrum. We shall see later that those requirements are naturally fulfilled by an oblique spacetime crystal, thus rendering a nonzero intrinsic current.

We consider a simple toy model in $(1+1)$D where an effective single-band Bloch system is subjected to a traveling wave potential. Since this is a toy model, we will not discuss its physical realization. Our start point will be the kernel matrix defined in Eq.~\eqref{HamiltonianKernel} with no freedom in index $\alpha/\beta$ given that the corresponding static Bloch system has only one band. We naively choose the matrix elements to be
\begin{equation}\label{caseStudyHamiltonian}
\begin{split}
    H_{m,n}(k) =& E\cos[(k+n\kappa)a]\delta_{m,n} + \Delta_1\delta_{m,n+1}+ \Delta_1^*\delta_{m,n-1} \\
    &+ \Delta_2\delta_{m,n+2}+ \Delta_2^*\delta_{m,n-2} - \hbar n\Omega \delta_{m,n},
\end{split}
\end{equation}
where $\Delta_{1,2}$ are the simplified hopping parameters in the frequency domain. Here, the corresponding static Bloch system has energy of form $E_B(k) = E\cos(k a)$ with $E>0$. We then consider a Fermi-Dirac distribution for the Bloch thermal equilibrium: $\rho^{B,eq}(
k) = \frac{1}{e^{[E_B(k)-\mu_C]/K_BT}+1}$ with $\mu_C$ being the chemical potential. 

\begin{figure}[t]
    \centering
    \includegraphics[width=0.5\textwidth]{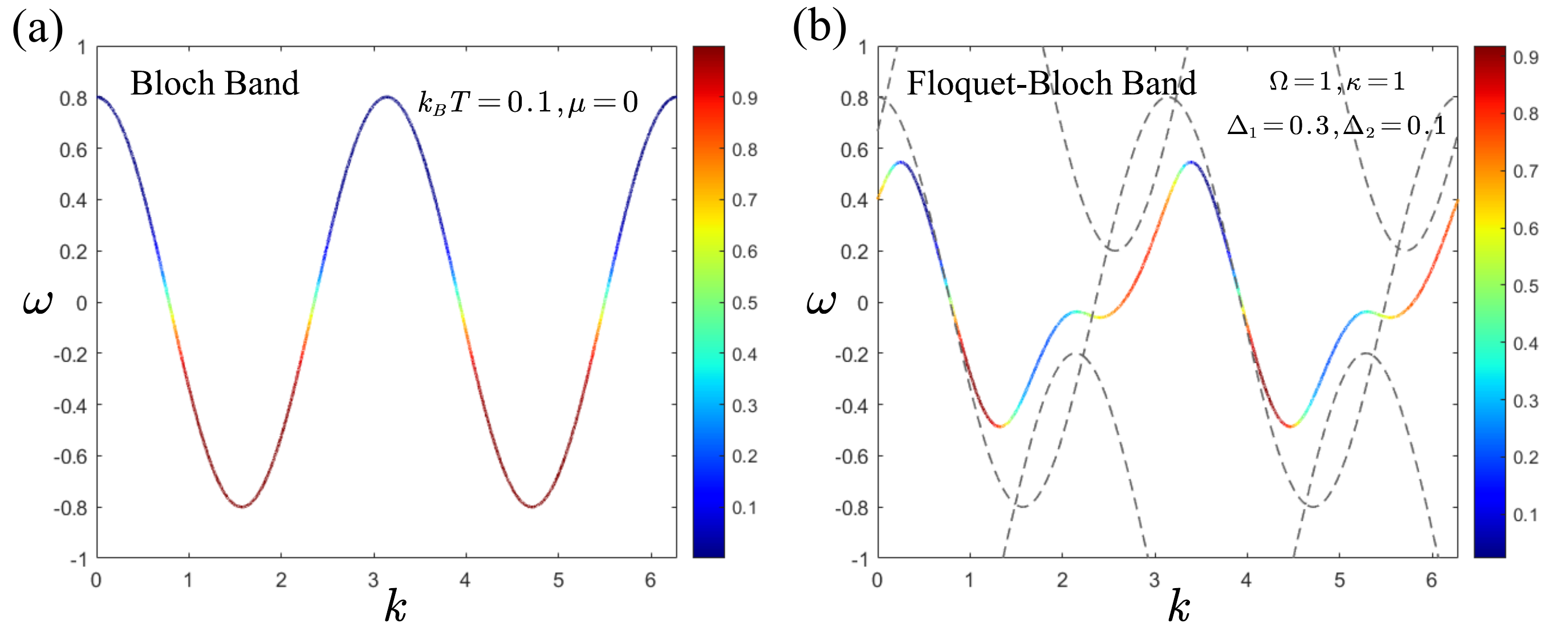}
    \caption{The electron populations in (a) the original Bloch band at the thermal equilibrium with $k_BT=0.1, \mu_C=0$ and (b) the Floquet-Bloch band after turning on the Floquet drive with $\Delta_1 = 0.3,\Delta_2=0.1$. Here $E_B = 0.8$ and $a = 2$. The color indicates the occupation number of each state. The dashed curves are the original Bloch band in (a) shifted by integer multiples of the phonon mode ($\Omega,\kappa$) in energy and momentum.}
    \label{electron_population}
\end{figure}

In Fig.~\ref{electron_population}, we plot the Bloch band and corresponding Floquet-Bloch band after turning on the Floquet drive, where the population of each state is indicated by different colors. One should see clearly that the Floquet-Bloch band is no longer symmetric and its electronic density at each state is no longer a function of the quasi-energy, since as we see that different $k$-states with the same quasi-energy have different colors (i.e. different populations). We claim that the inversion symmetry and the time-reversal symmetry are naturally broken in the oblique space-time crystal since a specific driving mode $(\Omega,\kappa)$ is chosen which selects unique directions in both space and time~\cite{Note5}. Thus, we expect a non-zero intrinsic current in an oblique spacetime crystal.

\begin{figure}[t]
    \centering
    \includegraphics[width=0.5\textwidth]{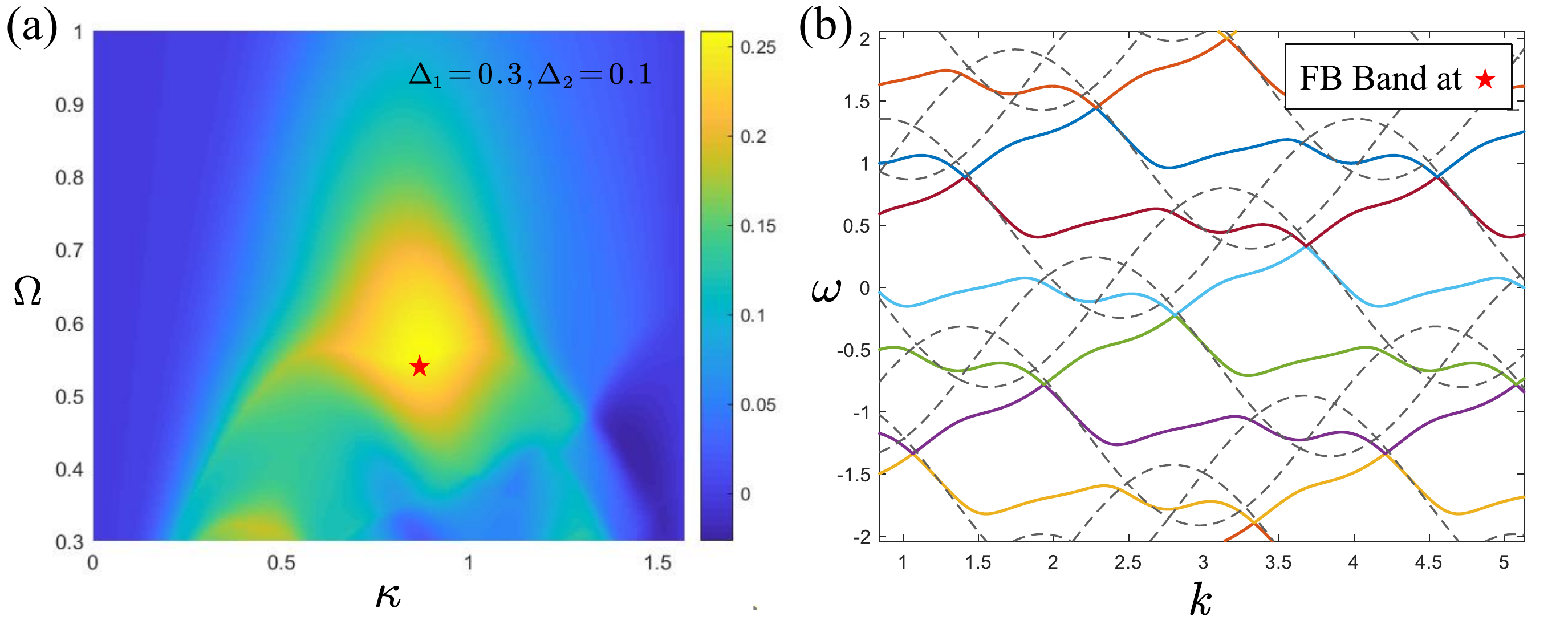}
    \caption{(a) The intrinsic current of the oblique spacetime crystal with different Floquet driving modes ($\Omega,\kappa$), where the specific point in such parameter space giving the maximal current is marked by a red star. The color indicates the value of the DC current achieved at each ($\Omega,\kappa$). (b) The Floquet-Bloch band structure corresponding to the red star in (a). The dashed curves are the original Bloch band shifted by integer multiples of the phonon mode ($\Omega,\kappa$) in energy and momentum. The other parameters used in this plot are the same as in Fig.~\ref{electron_population}.}
    \label{DC_current_plot}
\end{figure}

In Fig.~\ref{DC_current_plot}(a), the intrinsic current responses of the oblique spacetime at different driving modes ($\Omega,\kappa$) and fixed coupling strength are shown. It can be seen that there is a clear pattern for the current response in the parameter space. In Appendix~\ref{Floquet_gap}, we show that such a pattern resembles that of the Floquet direct band gap in the same parameter space, from which we can conclude that a smaller Floquet gap in general gives a higher intrinsic current. For example, we plot in Fig.~\ref{DC_current_plot}(b) the band structure of the Floquet-Bloch system corresponding to the parameters marked by the red star in Fig.~\ref{DC_current_plot}(a) that gives the maximal intrinsic current. One should notice that such intrinsic DC current is purely a consequence of the combination of the oblique Floquet drive ($\kappa\neq 0$) and the out-of-equilibrium electron populations ($\rho^F_{\mu,\mu} $ is not a function of quasi-energy). In some cases, such oblique Floquet drive can be realized by coherent acoustic waves~\cite{gao2021floquet}, therefore, the intrinsic current can also be concluded as an acoustoelectric effect but in the strongly driven regime. 

\section{Conclusion and Remarks}\label{discussion}
In the first half of this work, we developed a semiclassical theory for the electrons in space-time crystals or Floquet-Bloch systems. The equations of motion of a single Floquet-Bloch electron under external slowly varying fields are derived, which shows similar behavior as the ordinary Bloch electrons but with modified quantities in the Floquet context. We then discussed the response of the Floquet-Bloch system to electromagnetic fields and found the intrinsic magnetization of the wave-packet due to its self-rotation in a magnetic field. Furthermore, we looked at the polarization of a fully occupied Floquet band which has two contributions: one is the original Bloch polarization modified by the Floquet coefficients; the other comes purely from the time variations. The latter contribution can be shown to vanish exactly under the $\mathcal{PT}$ symmetry. It is worth noting that a totally different approach to constructing the semiclassical theory for Floquet-Bloch electrons can be found in Ref.~\cite{dong2021geometrodynamics}, which gives qualitatively the same results. In the second half, we investigated the density matrix of the Floquet-Bloch system attached to a heat bath, which shows the essential distinction between a periodically driven Bloch system and a static one. The density matrix is shown to be generally non-diagonal and time-dependent for nonzero damping, while it becomes diagonal and time-independent when the damping is turned off. By using the diagonal density matrix, we can calculate the intrinsic DC current, which is nonzero indicating the non-equilibrium nature of the Floquet-Bloch system. We further discussed the implications when finite damping is included and found that the semiclassical scheme developed in this work cannot be applied to the damping Floquet-Bloch system. Further developments are certainly needed to overcome this inapplicability. In the rest of this section, a few remarks are given emphasizing the potential of this semiclassical method combined with the Floquet-Bloch systems.

Firstly, we want to comment on the non-equilibrium nature of the Floquet-Bloch system which leads to strangely behaved electron densities at different eigenstates. As we have already mentioned in Sec.~\ref{caseStudy}, the electron populations in the steady state are no longer a function of the quasi-energy. For example, the relation $\frac{\partial \rho}{\partial \bk}= \frac{\partial \rho}{\partial \omega}\frac{\partial \omega}{\partial \bk}$ does not hold in general. In contrast to the equilibrium electron distribution that cares only about the energy of the electron, the non-equilibrium distribution discussed in this work has also strong momentum dependence. A direct consequence of such momentum dependence is the intrinsic non-zero DC current. We expect more physical implications of such non-energy-dependent distribution to occur in higher dimensions.

Secondly, we emphasize that the semiclassical theory developed and applied in this work only deals with DC responses (slow time variations) of the Floquet-Bloch systems, given that the oscillating part (fast time variations) has been averaged. However, in periodically driven systems, there are still physical observables sensitive to the fast time variations, such as the higher harmonic generations in strong light-matter interactions~\cite{macklin1993high}. Even for DC responses, sometimes there could be cases where we can get contributions from the high harmonics, as one can see from the fact that $\langle\langle \rho \hat{O} \rangle\rangle\neq \langle\langle \rho \rangle\rangle \langle\langle \hat{O} \rangle\rangle$ if $\rho$ is non-diagonal and/or contains fast time variations. In those cases (for example, systems with finite damping), careful evaluations of $\langle\langle \rho \hat{O} \rangle\rangle$ are needed.

Lastly, in this work, our scope is mostly limited to one spatial dimension in the applications, where only a few physics can be studied. Moreover, the spin degrees of freedom and other possible internal degrees of freedom such as the valley or pseudo-spins are completely ignored in our consideration. We expect that, in future developments, the semiclassical formalism can be used to discuss more exotic physics with more degrees of freedom and in higher dimensions, for example, the spin-polarized photo-currents~\cite{berdakin2021spin}, the Hall effect in periodically driven systems~\cite{dehghani2015out}, the light-induced magnetization~\cite{chovan2006ultrafast}, and/or the electronic transport in Floquet topological insulators~\cite{lindner2011floquet,rudner2020band}.

\appendix
\section{Orthonormal condition for the Floquet-Bloch wavefunction}\label{orthonormalCondition}
Given the Floquet-Bloch eigen-wavefunction:
\begin{equation}
    |\Psi^\mu_{\boldsymbol{k}}(\boldsymbol{r},t)\rangle = \sum_{n,\beta}f^{\mu,\beta}_{n,\boldsymbol{k}}e^{-in\Omega t}|\varphi^\alpha_{\boldsymbol{k}+n\boldsymbol{\kappa}}(\boldsymbol{r})\rangle,
\end{equation}
we have
\begin{align}
        &\langle \Psi^\mu_{\boldsymbol{k}}(\boldsymbol{r},t)|\Psi^\nu_{\boldsymbol{k}'}(\boldsymbol{r},t)\rangle \notag\\
        &= \sum_{n,m}\sum_{\alpha,\beta}\left( f^{\mu,\alpha}_{n,\bk} \right)^*e^{in\Omega t}\langle\varphi^\alpha_{\boldsymbol{k}+n\boldsymbol{\kappa}}(\boldsymbol{r})|\varphi^\beta_{\bk'+m\boldsymbol{\kappa}}(\boldsymbol{r}) \rangle e^{-im\Omega t}f^{\nu,\beta}_{m,\bk'} \notag\\
        &=\sum_{\bG}\sum_{n,m}\sum_{\alpha,\beta}e^{i(n-m)\Omega t}\left( f^{\mu,\alpha}_{n,\bk} \right)^*f^{\nu,\beta}_{m,\bk'}\notag\\
        &\quad\times\delta_{\alpha,\beta}\delta(\bk+n\boldsymbol{\kappa}-\bk'-m\boldsymbol{\kappa}+\bG) \notag\\
        &= \sum_{\bG}\sum_{q}e^{-i q\Omega t}\left[\sum_{n,\alpha}\left( f^{\mu,\alpha}_{n,\bk} \right)^*f^{\nu,\alpha}_{n+q,\bk-q\boldsymbol{\kappa}+\bG}\right]\notag\\
        &\quad\times\delta(\bk-\bk'-q\boldsymbol{\kappa}+\bG) \notag\\
        &= \sum_{\bG}\sum_{q}e^{-i q\Omega t}\delta_{\mu,\nu}\delta_{q,0}\delta(\bk-\bk'-q\boldsymbol{\kappa}+\bG) \notag\\
        &= \sum_{\bG}\delta_{\mu,\nu}\delta(\bk-\bk'+\bG),
\end{align}
where we have used the relation in Eq.~\eqref{orthogonality}. Here $\bG$ is the reciprocal lattice vector of the corresponding static Bloch crystal. It is worthwhile to note that the orthonormal condition of the Floquet-Bloch eigen-wavefunction does not involve the time integral.

\section{The position center of the wave-packet}\label{positioncenter}
The position center of a Floquet-Bloch wave-packet is given by the expectation value of the position operator evaluated under the wave-packet wavefunction
\begin{align}
        \boldsymbol{r}_c(\tau_s) \equiv&\langle\langle W_\mu|\hat{\boldsymbol{r}}|W_\mu\rangle\rangle 
        = \sum_{nn',\alpha\beta}\frac{1}{T}\int_{\tau_s}^{\tau_s+T}d\tau_f e^{-i(n'-n)\Omega \tau_f} \notag\\ 
        &\times\int d\boldsymbol{k}d\boldsymbol{k}'|a(\boldsymbol{k},\tau_s)|e^{i\gamma(\boldsymbol{k},\tau_s)}\left( f^{\mu,\alpha}_{n,\boldsymbol{k}} \right)^*\notag\\
        &\times\langle\varphi_{n,\boldsymbol{k}}^{\alpha}|\hat{\boldsymbol{r}}|\varphi_{n',\boldsymbol{k}'}^{\beta}\rangle f^{\mu,\beta}_{n',\boldsymbol{k}'}e^{-i\gamma(\boldsymbol{k}',\tau_s)}|a(\boldsymbol{k}',\tau_s)|,
\end{align}
where the position operator $\hat{\boldsymbol{r}}$ sandwiched by the Bloch states can be evaluated as~\cite{sundaram1999wave,blount1962formalisms}
\begin{align}\label{positionInBloch}            &\langle\varphi_{n,\boldsymbol{k}}^{\alpha}|\hat{\boldsymbol{r}}|\varphi_{n',\boldsymbol{k}'}^{\beta}\rangle \notag\\
    &= \left[ i\partial_{\bk_n}\delta_{\alpha,\beta} + \langle u_{n,\boldsymbol{k}}^{\alpha}|i\partial_{\bk'_{n'}}|u_{n',\boldsymbol{k}'}^{\beta}\rangle \right]\delta(\bk_n-\bk'_{n'}),
\end{align}
with $\bk_n \equiv \bk + n\boldsymbol\kappa$. Notice that we can first integrate out the fast time dependence part: $\frac{1}{T}\int_{\tau_s}^{\tau_s+T}e^{-i(n'-n)\Omega \tau_f} d\tau_f = \delta_{n,n'}$, thus the expression for the position center becomes
\begin{align}
        &\boldsymbol{r}_c(\tau_s)
       =  \sum_{n,\alpha\beta} \int d\boldsymbol{k}\, |a(\boldsymbol{k},\tau_s)|^2 \left\{ \delta_{\alpha,\beta}\left[ \left( f^{\mu,\alpha}_{n,\boldsymbol{k}} \right)^*f^{\mu,\beta}_{n,\boldsymbol{k}}\partial_{\bk}\gamma(\boldsymbol{k}',\tau_s)\right.\right.\notag\\
        &\quad+\left.\left.i \left( f^{\mu,\alpha}_{n,\boldsymbol{k}} \right)^*\partial_{\bk} f^{\mu,\beta}_{n,\boldsymbol{k}}\right]  + \langle u_{n,\boldsymbol{k}}^{\alpha}|i\partial_{\bk'_n}|u_{n,\boldsymbol{k}'}^{\beta}\rangle \left(f^{\mu,\alpha}_{n,\boldsymbol{k}} \right)^*f^{\mu,\beta}_{n,\boldsymbol{k}} \right\} \notag\\
        &\quad+ \sum_{n,\alpha} \int d\boldsymbol{k}\,|f^{\mu,\alpha}_{n,\boldsymbol{k}} |^2 |a(\boldsymbol{k},\tau_s)|\partial_{\bk} |a(\boldsymbol{k},\tau_s)|.
\end{align}
The last term actually vanishes given that $ \sum_{n,\alpha}|f^{\mu,\alpha}_{n,\boldsymbol{k}} |^2 = 1$ and $\int d\boldsymbol{k}\,\partial_{\bk}|a(\boldsymbol{k},\tau_s)|^2 = 0 $,
which follows from the fact that $|a(\boldsymbol{k},\tau_s)|^2\sim \delta(\bk-\bk_c(\tau_s)) $. And we also have for an arbitrary function $f$ depending slowly on the quasi-momentum $\bk$: $\int d\bk |a(\bk,\tau_s)|^2f(\bk)\approx f(\bk_c(\tau_s))$. So, the position center can be evaluated as
\begin{align}\label{positionCenter}
    &\boldsymbol{r}_c(\tau_s) \approx  \partial_{\boldsymbol{k}_c}\gamma(\boldsymbol{k}_c,\tau_s) \notag\\
    &\quad+ \sum_{n,\alpha}\left[ i(f^{\mu,\alpha}_{n,\boldsymbol{k}_c})^*\partial_{\boldsymbol{k}_c}f^{\mu,\alpha}_{n,\boldsymbol{k}_c} + \sum_{\beta}\mathcal{A}_{n,\boldsymbol{k}_c}^{\alpha,\beta} (f^{\mu,\alpha}_{n,\boldsymbol{k}_c})^*f^{\mu,\beta}_{n,\boldsymbol{k}_c}\right]\notag\\
    &\equiv \partial_{\boldsymbol{k}_c}\gamma(\boldsymbol{k}_c,\tau_s) + \left.\langle\langle \Tilde{u}_{\boldsymbol{k}}^{\mu}|i\partial_{\boldsymbol{k}}| \Tilde{u}_{\boldsymbol{k}}^{\mu} \rangle\rangle\right|_{\boldsymbol{k}=\boldsymbol{k}_c}
\end{align}
where $\mathcal{A}_{n,\boldsymbol{k}_c}^{\alpha,\beta}$ is the interband Berry connection at $\boldsymbol{k}_c+n\boldsymbol{\kappa}$ of the corresponding static Bloch system without Floquet drive, and $ | \Tilde{u}_{\boldsymbol{k}}^{\mu} \rangle = \sum_n f^{\mu,\alpha}_{n,\bk}e^{-in\Omega \tau_f}e^{in\boldsymbol{\kappa}\cdot\br}|u^\alpha_{n,\bk}\rangle $ is the periodic part of the Floquet-Bloch wave-function.

The position center $\br_c$ in the Floquet context has a special meaning. The fast variations that have been averaged over the Floquet period correspond to oscillating motions of the electron around $\br_c$. Thus, one can view $\br_c$ as the guiding center of the electron, which then can carry DC current.

\section{The time derivative of the wave-packet}\label{time_derivative_wavepacket}
Next, let us look at the time derivative of the wave-packet, which is the expectation value of the operator $id/dt$:
\begin{align}
        &\langle\langle W_\mu |i\frac{d}{d t} | W_\mu \rangle\rangle =\sum_{n n',\alpha\beta}\frac{1}{T}\int_{\tau_s}^{\tau_s+T}d \tau_f e^{in\Omega \tau_f}\int d\boldsymbol{k}d\boldsymbol{k}' \notag\\
        &\times |a(\boldsymbol{k},\tau_s)|e^{i\gamma(\boldsymbol{k},\tau_s)}\left( f^{\mu,\alpha}_{n,\boldsymbol{k}} \right)^*\langle\varphi_{n,\boldsymbol{k}}^{\alpha}|i\left(\frac{d}{d \tau_f} + \frac{d}{d \tau_s}\right)\left[|\varphi_{n',\boldsymbol{k}'}^{\beta}\rangle\right.\notag\\
        &\times \left.f^{\mu,\beta}_{n',\boldsymbol{k}'}e^{-i\gamma(\boldsymbol{k}',\tau_s)}|a(\boldsymbol{k}',\tau_s)|e^{-in'\Omega \tau_f}\right],
\end{align}
where we explicitly separate the contributions between slowly varying and fast varying parts. As we can see, the only fast varying part is the time-dependent phase factors $e^{in\Omega \tau_f}$ and $e^{-in'\Omega \tau_f}$, so the fast time dependence can be readily integrated out:
\begin{align}
        &\langle\langle W_\mu |i\frac{d}{d t} | W_\mu \rangle\rangle \notag\\
        &= \sum_{n,\alpha\beta}\int d\boldsymbol{k}d\boldsymbol{k}'  |a(\boldsymbol{k},\tau_s)| e^{i\gamma(\boldsymbol{k},\tau_s)}\left( f^{\mu,\alpha}_{n,\boldsymbol{k}} \right)^*\langle\varphi_{n,\boldsymbol{k}}^{\alpha}|\notag\\
        &\qquad\times\left(n\Omega + i\frac{d}{d \tau_s}\right)|\varphi_{n,\boldsymbol{k}'}^{\beta}\rangle f^{\mu,\beta}_{n,\boldsymbol{k}'}e^{-i\gamma(\boldsymbol{k}',\tau_s)}|a(\boldsymbol{k}',\tau_s)|.
\end{align}
Now, we should be careful when counting the slow time dependence of the wave function. First of all, we have explicit slow time dependence in $|a(\bk,\tau_s)|$ and $|\gamma(\bk,\tau_s)|$. The coefficient $f^{\mu,\beta}_{n,\boldsymbol{k}}(\br_c)$ depending on $\bk$ and the position center $\br_c$ has slow time dependence through $\br_c(\tau_s)$ and will get another time dependence after taking the $\bk$-integration (getting it from the weight $|a(\bk,\tau_s)|$). Then the most tricky one is the slow time dependence of the Bloch wave-function $ |\varphi_{n,\boldsymbol{k}}^{\alpha}\rangle$ which is obtained through the position center dependence: $ |\varphi_{n,\boldsymbol{k}}^{\alpha}\rangle\to|\varphi_{n,\boldsymbol{k}}^{\alpha}(\br_c)\rangle = e^{i(\bk+n\boldsymbol{\kappa})\cdot\hat{\br}}|u_{n,\boldsymbol{k}}^{\alpha}(\br_c)\rangle$. Having this extra dependence, we can write $ d_{\tau_s} = \partial_{\tau_s} + \dot{\br}_c\cdot\partial_{\br_c}$, where $\dot\bO\equiv d\bO/d\tau_s$ means slow time derivative. Similarly, we have
\begin{align}\label{timeVary}
&\langle\varphi_{n,\boldsymbol{k}}^{\alpha}|i\frac{d}{d \tau_s}|\varphi_{n',\boldsymbol{k}'}^{\beta}\rangle \notag\\
&= \left[ i\frac{d}{d \tau_s}\delta_{\alpha,\beta} + \langle u_{n,\boldsymbol{k}}^{\alpha}|i\frac{d}{d \tau_s}|u_{n,\boldsymbol{k}}^{\beta}\rangle \right]\delta(\bk_n-\bk'_{n'}),
\end{align}
Therefore, by plugging Eq.~\eqref{timeVary} into the time evolution, we get
\begin{align}
        &\langle\langle W_\mu |i\frac{d}{d t} | W_\mu \rangle\rangle =\partial_{\tau_s}\gamma(\bk_c,\tau_s)+  \sum_{n,\alpha}\left| f^{\mu,\alpha}_{n,\boldsymbol{k}_c}\right|^2n\Omega \notag\\
        &\quad+ \left.\langle\langle \Tilde{u}_{\boldsymbol{k}}^{\mu}|i\partial_{\tau_s}| \Tilde{u}_{\boldsymbol{k}}^{\mu} \rangle\rangle\right|_{\boldsymbol{k}=\boldsymbol{k}_c} +\Dot{\boldsymbol{r}}_c\cdot\left.\langle\langle \Tilde{u}_{\boldsymbol{k}}^{\mu}|i\partial_{\boldsymbol{r}_c}| \Tilde{u}_{\boldsymbol{k}}^{\mu} \rangle\rangle\right|_{\boldsymbol{k}=\boldsymbol{k}_c},
\end{align}
where we have used the fact that $\partial_{\tau_s}\int d\boldsymbol{k}\, |a(\boldsymbol{k},\tau_s)|^2 =0$.
We want to emphasize here that there are three different length scales: the lattice constant $a$ of the static crystal; the wavelength of the propagating wave $2\pi/|\kappa|$; and the typical length of the external field $L$. The change of $\br_c$ should only track the variation of the external field, thus we may require a separation of the external length scale from the other two scales. Typically, it will be $L\gg 2\pi/|\kappa|\sim a$ or $2\pi/|\kappa|\gg L\gg  a$.
Notice that after the momentum integration, all $\bk$-dependent quantities will become $\bk_c(\tau_s)$-dependent inherited from the weight function $|a(\bk,\tau_s)|^2$, which then introduces extra slow time dependence. It is important to observe that
\begin{align}
    \partial_{\tau_s}\gamma(\bk_c,\tau_s) =& \dot\gamma(\bk_c,\tau_s) - \dot\bk_c\cdot\partial_{\bk_c}\gamma(\bk_c,\tau_s) \notag\\
    =& \dot\gamma(\bk_c,\tau_s) - \dot\bk_c\cdot\left[\br_c- \left.\langle\langle \Tilde{u}_{\boldsymbol{k}}^{\mu}|i\partial_{\boldsymbol{k}}| \Tilde{u}_{\boldsymbol{k}}^{\mu} \rangle\rangle\right|_{\boldsymbol{k}=\boldsymbol{k}_c} \right],
\end{align}
based on which, we have
\begin{align}\label{timeDerivative}
        &\langle\langle W_\mu |i\frac{d}{d t} | W_\mu \rangle\rangle =\dot\gamma(\bk_c,\tau_s) \notag\\
        &- \dot\bk_c\cdot\br_c+  \sum_{n,\alpha}\left| f^{\mu,\alpha}_{n,\boldsymbol{k}_c}\right|^2n\Omega + \left.\langle\langle \Tilde{u}_{\boldsymbol{k}}^{\mu}|i\partial_{\tau_s}| \Tilde{u}_{\boldsymbol{k}}^{\mu} \rangle\rangle\right|_{\boldsymbol{k}=\boldsymbol{k}_c} \notag\\
        &+ \Dot{\boldsymbol{r}}_c\cdot\left.\langle\langle \Tilde{u}_{\boldsymbol{k}}^{\mu}|i\partial_{\boldsymbol{r}_c}| \Tilde{u}_{\boldsymbol{k}}^{\mu} \rangle\rangle\right|_{\boldsymbol{k}=\boldsymbol{k}_c}+\dot\bk_c\cdot\left.\langle\langle \Tilde{u}_{\boldsymbol{k}}^{\mu}|i\partial_{\boldsymbol{k}}| \Tilde{u}_{\boldsymbol{k}}^{\mu} \rangle\rangle\right|_{\boldsymbol{k}=\boldsymbol{k}_c}.
\end{align}
It is important to note that there are no fast-time derivatives in the above equation since the fast-time variations have been integrated by taking the time average over the Floquet period. This is again based on the assumption of two separate time scales.

\section{Energy of the wave-packet and its gradient correction}\label{energyCorrectionAppendix}
The wave-packet constructed is essentially an extended object with finite size in both real space and momentum space. Thus, the potential energy of such an extended object will have gradient corrections additional to the energy simply evaluated at the position center $\br_c$.
The energy of the wave-packet can be evaluated as
\begin{align}\label{averagedEnergy}
    \mathcal{E}_\mu &\equiv \langle\langle W_\mu | \hat{H}(\hat{\boldsymbol{r}},t)| W_\mu \rangle\rangle \notag\\
    &\approx \langle\langle W_\mu | \hat{H}_c(\hat{\boldsymbol{r}},\tau_f;\boldsymbol{r}_c,\tau_s)| W_\mu \rangle\rangle + \langle\langle W_\mu | \Delta\hat{H}| W_\mu \rangle\rangle \notag\\
    &\equiv \mathcal{E}^0_{\mu}(\boldsymbol{r}_c,\boldsymbol{k}_c,\tau_s) + \Delta\mathcal{E}_\mu,
\end{align}
where the first order gradient correction to the Hamiltonian reads
\begin{equation}
    \Delta\hat{H} = \frac{1}{2}\left[ (\hat{\boldsymbol{r}}-\boldsymbol{r}_c)\cdot\frac{\partial\hat{H}_c}{\partial\boldsymbol{r}_c} + \frac{\partial\hat{H}_c}{\partial\boldsymbol{r}_c}\cdot (\hat{\boldsymbol{r}}-\boldsymbol{r}_c)\right].
\end{equation}
It is easy to show that the leading order reads
\begin{equation}\label{leadingOrderEnergy}
    \begin{split}
        &\mathcal{E}^0_{\mu}(\boldsymbol{r}_c,\boldsymbol{k}_c,\tau_s)= \hbar\omega_\mu(\bk_c,\br_c)+\hbar\sum_{n,\alpha} n\Omega\left|f^{\mu,\alpha}_{n,\bk_c}\right|^2.
    \end{split}
\end{equation}
Then the gradient correction to the energy is
\begin{align}
        \Delta\mathcal{E}_\mu &= \frac{1}{2}\langle\langle W_\mu | (\hat{\boldsymbol{r}}-\boldsymbol{r}_c)\cdot\frac{\partial\hat{H}_c}{\partial\boldsymbol{r}_c} + \frac{\partial\hat{H}_c}{\partial\boldsymbol{r}_c}\cdot (\hat{\boldsymbol{r}}-\boldsymbol{r}_c)| W_\mu \rangle\rangle \notag\\
        &= \Re\,\langle\langle W_\mu | \frac{\partial\hat{H}_c}{\partial\boldsymbol{r}_c}\cdot(\hat{\boldsymbol{r}}-\boldsymbol{r}_c)| W_\mu \rangle\rangle.
\end{align}

\begin{widetext}
Evaluating the gradient correction requires some algebra which will be performed below.
Let us first evaluate the following quantity
\begin{align}\label{relation1}
    &\sum_{nn',\alpha\beta}\left( f^{\mu,\alpha}_{n,\boldsymbol{k}} \right)^*\langle  \Phi^\alpha_{n,\bk}(\br,\br_c)|\frac{\partial\hat{H}_c}{\partial\boldsymbol{r}_c}|\Phi^{\beta}_{n',\bk'}(\br,\br_c)\rangle f^{\nu,\beta}_{n',\boldsymbol{k}'} \notag\\
    &= \frac{\partial}{\partial\boldsymbol{r}_c}\left[\sum_{nn',\alpha\beta}\left( f^{\mu,\alpha}_{n,\boldsymbol{k}} \right)^*\langle \Phi^\alpha_{n,\bk}(\br,\br_c)|\hat{H}_c|\Phi^{\beta}_{n',\bk'}(\br,\br_c)\rangle f^{\nu,\beta}_{n',\boldsymbol{k}'}\right] -  \sum_{nn',\alpha\beta}\frac{\partial}{\partial\boldsymbol{r}_c}\left[\left( f^{\mu,\alpha}_{n,\boldsymbol{k}} \right)^*\langle  \Phi^\alpha_{n,\bk}(\br,\br_c)|\right]\hat{H}_c|\Phi^{\beta}_{n',\bk'}(\br,\br_c)\rangle f^{\nu,\beta}_{n',\boldsymbol{k}'} \notag\\
    &\qquad - \sum_{nn',\alpha\beta}\left( f^{\mu,\alpha}_{n,\boldsymbol{k}} \right)^*\langle  \Phi^\alpha_{n,\bk}(\br,\br_c)|\hat{H}_c\frac{\partial}{\partial\boldsymbol{r}_c}\left[|\Phi^{\beta}_{n',\bk'}(\br,\br_c)\rangle f^{\nu,\beta}_{n',\boldsymbol{k}'}\right] \notag\\
    &= \hbar\frac{\partial\omega_\mu(\bk,\br_c)}{\partial\br_c}\delta_{\mu,\nu}\delta(\bk-\bk') - \sum_{nn',\alpha\beta}\hbar[\omega_{\nu}(\bk',\br_c)-\omega_\mu(\bk,\br_c)+(n'-n)\Omega]\frac{\partial}{\partial\boldsymbol{r}_c}\left[\left( f^{\mu,\alpha}_{n,\boldsymbol{k}} \right)^*\langle  \Phi^\alpha_{n,\bk}(\br,\br_c)|\right]|\Phi^{\beta}_{n',\bk'}(\br,\br_c)\rangle f^{\nu,\beta}_{n',\boldsymbol{k}'}
\end{align}
where we have utilized facts that
\begin{equation}
\hat{H}_c\sum_{n,\alpha}f^{\mu,\alpha}_{n,\boldsymbol{k}}|\Phi^{\alpha}_{n,\bk}(\br,\br_c)\rangle =\sum_{n,\alpha}(\omega_\mu(\bk,\br_c)+n\Omega)f^{\mu,\alpha}_{n,\boldsymbol{k}}|\Phi^{\alpha}_{n,\bk}(\br,\br_c)\rangle .
\end{equation}
Now, we can proceed to evaluate the energy correction. We have
\begin{align}
        &\langle\langle W_\mu | \frac{\partial\hat{H}_c}{\partial\boldsymbol{r}_c}\cdot\br_c| W_\mu \rangle\rangle = \br_c\cdot\langle\langle W_\mu | \frac{\partial\hat{H}_c}{\partial\boldsymbol{r}_c}| W_\mu \rangle\rangle \notag\\
        &= \br_c\cdot \sum_{nn',\alpha\beta} \int d\boldsymbol{k}d\boldsymbol{k}'\, |a(\boldsymbol{k},\tau_s)|e^{i\gamma(\boldsymbol{k},\tau_s)}\left( f^{\mu,\alpha}_{n,\boldsymbol{k}} \right)^*\langle\langle  \Phi^\alpha_{n,\bk}(\br,\br_c)|\frac{\partial\hat{H}_c}{\partial\boldsymbol{r}_c}|\Phi^{\beta}_{n',\bk'}(\br,\br_c)\rangle\rangle f^{\mu,\beta}_{n',\boldsymbol{k}'}e^{-i\gamma(\boldsymbol{k}',\tau_s)}|a(\boldsymbol{k}',\tau_s)|\notag\\
        &= \br_c\cdot\frac{\partial\hbar\omega_\mu(\bk_c,\br_c)}{\partial\br_c},
\end{align}
where we have used Eq.~\eqref{relation1} and assumed that the spatial variation of the external field is small so that the momentum is still preserved, namely
\begin{equation}\label{MomentumConservation}
    \frac{\partial}{\partial\boldsymbol{r}_c}\left[\left( f^{\mu,\alpha}_{n,\boldsymbol{k}} \right)^*\langle  \Phi^\alpha_{n,\bk}(\br,\br_c)|\right]|\Phi^{\beta}_{n',\bk'}(\br,\br_c)\rangle f^{\nu,\beta}_{n',\boldsymbol{k}'} = \frac{\partial}{\partial\boldsymbol{r}_c}\left[\left( f^{\mu,\alpha}_{n,\boldsymbol{k}} \right)^*\langle  \Phi^\alpha_{n,\bk}(\br,\br_c)|\right]|\Phi^{\beta}_{n',\bk'}(\br,\br_c)\rangle f^{\nu,\beta}_{n',\boldsymbol{k}'}\delta(\bk_n-\bk'_{n'}).
\end{equation}
Then,
\begin{align}
        \langle\langle W_\mu | \frac{\partial\hat{H}_c}{\partial\boldsymbol{r}_c}\cdot\hat{\boldsymbol{r}}| W_\mu \rangle\rangle 
         &= \sum_{nn',\alpha\beta}\sum_{\nu}\sum_{\alpha'\beta',mm'} \int d\boldsymbol{k}d\boldsymbol{k}'d\bk''\, |a(\boldsymbol{k},\tau_s)|e^{i\gamma(\boldsymbol{k},\tau_s)}\left( f^{\mu,\alpha}_{n,\boldsymbol{k}} \right)^*\Big\langle\langle  \Phi^\alpha_{n,\bk}(\br,\br_c)|\frac{\partial\hat{H}_c}{\partial\boldsymbol{r}_c}|\Phi^{\alpha'}_{m,\bk''}(\br,\br_c)\rangle f^{\nu,\alpha'}_{m,\boldsymbol{k}''}\notag\\
        &\qquad\cdot\left( f^{\nu,\beta'}_{m',\boldsymbol{k}''} \right)^*\langle  \Phi^{\beta'}_{m',\bk''}(\br,\br_c)|\hat{\boldsymbol{r}}|\Phi^{\beta}_{n',\bk'}(\br,\br_c)\rangle\Big\rangle f^{\mu,\beta}_{n',\boldsymbol{k}'}e^{-i\gamma(\boldsymbol{k}',\tau_s)}|a(\boldsymbol{k}',\tau_s)|,
\end{align}
where we have inserted an identity operator from the completeness of the Floquet-Bloch basis:
\begin{equation}
    \mathcal{I} = \sum_{\nu}\sum_{\alpha'\beta',mm'}\int d\bk'' |\Phi^{\alpha'}_{m,\bk''}(\br,\br_c)\rangle f^{\nu,\alpha'}_{m,\boldsymbol{k}''} \left( f^{\nu,\beta'}_{m',\boldsymbol{k}''} \right)^*\langle  \Phi^{\beta'}_{m',\bk''}(\br,\br_c)|.
\end{equation}
Again, using Eq.~\eqref{relation1}, we have
\begin{align}
        &\langle\langle W_\mu | \frac{\partial\hat{H}_c}{\partial\boldsymbol{r}_c}\cdot\hat{\boldsymbol{r}}| W_\mu \rangle\rangle = \Big\langle\hbar\sum_{n',\beta}\sum_{\nu}\sum_{\beta',m'} \int d\boldsymbol{k}d\boldsymbol{k}'d\bk''\, |a(\boldsymbol{k},\tau_s)|e^{i\gamma(\boldsymbol{k},\tau_s)}\Bigg\{\frac{\partial\omega_\mu(\bk,\br_c)}{\partial\br_c}\delta_{\mu,\nu}\delta(\bk-\bk'')\notag\\
        &\quad -\sum_{nm,\alpha\alpha'}[\omega_{\nu}(\bk'',\br_c)-\omega_\mu(\bk,\br_c)+(m-n)\Omega]\frac{\partial}{\partial\boldsymbol{r}_c}\left[\left( f^{\mu,\alpha}_{n,\boldsymbol{k}} \right)^*\langle  \Phi^\alpha_{n,\bk}(\br,\br_c)|\right]|\Phi^{\alpha'}_{m,\bk''}(\br,\br_c)\rangle f^{\nu,\alpha'}_{m,\bk''} \Bigg\} \notag\\
        &\quad\cdot\left( f^{\nu,\beta'}_{m',\boldsymbol{k}''} \right)^*\langle  \Phi^{\beta'}_{m',\bk''}(\br,\br_c)|\hat{\boldsymbol{r}}|\Phi^{\beta}_{n',\bk'}(\br,\br_c)\rangle f^{\mu,\beta}_{n',\boldsymbol{k}'}e^{-i\gamma(\boldsymbol{k}',\tau_s)}|a(\boldsymbol{k}',\tau_s)|\Big\rangle,
\end{align}
which contains two terms. The first term is
\begin{align}
        &\hbar\sum_{n',\beta}\sum_{\nu}\sum_{\beta',m'} \int d\boldsymbol{k}d\boldsymbol{k}'d\bk''\, |a(\boldsymbol{k},\tau_s)|e^{i\gamma(\boldsymbol{k},\tau_s)}\delta_{\mu,\nu}\delta(\bk-\bk'') \notag\\
        &\qquad\times\frac{\partial\omega_\mu(\bk,\br_c)}{\partial\br_c}\cdot\left( f^{\nu,\beta'}_{m',\boldsymbol{k}''} \right)^*\langle\langle  \Phi^{\beta'}_{m',\bk''}(\br,\br_c)|\hat{\boldsymbol{r}}|\Phi^{\beta}_{n',\bk'}(\br,\br_c)\rangle\rangle f^{\mu,\beta}_{n',\boldsymbol{k}'}e^{-i\gamma(\boldsymbol{k}',\tau_s)}|a(\boldsymbol{k}',\tau_s)|\notag\\
        &=\hbar\sum_{n',\beta}\sum_{\beta',m'} \int d\boldsymbol{k}d\boldsymbol{k}'\, |a(\boldsymbol{k},\tau_s)|e^{i\gamma(\boldsymbol{k},\tau_s)}\frac{\partial\omega_\mu(\bk,\br_c)}{\partial\br_c}\cdot\left( f^{\mu,\beta'}_{m',\boldsymbol{k}''} \right)^*\langle\langle  \Phi^{\beta'}_{m',\bk''}(\br,\br_c)|\hat{\boldsymbol{r}}|\Phi^{\beta}_{n',\bk'}(\br,\br_c)\rangle\rangle f^{\mu,\beta}_{n',\boldsymbol{k}'}e^{-i\gamma(\boldsymbol{k}',\tau_s)}|a(\boldsymbol{k}',\tau_s)|\notag\\
        &=\frac{\partial\hbar\omega_\mu(\bk_c,\br_c)}{\partial\br_c}\cdot\br_c,
\end{align}
where we have used the definition for the position center $\br_c$.
The second term is much more complicated, which can be calculated as
\begin{align}\label{secondTerm}
        &\Big\langle\hbar\sum_{nn',\alpha\beta}\sum_{\nu}\sum_{\alpha'\beta',mm'} \int d\boldsymbol{k}d\boldsymbol{k}'d\bk''\, |a(\boldsymbol{k},\tau_s)|e^{i\gamma(\boldsymbol{k},\tau_s)}[\omega_\mu(\bk,\br_c)-\omega_{\nu}(\bk'',\br_c)+(n-m)\Omega]\frac{\partial}{\partial\boldsymbol{r}_c}\left[\left( f^{\mu,\alpha}_{n,\boldsymbol{k}} \right)^*\langle  \Phi^\alpha_{n,\bk}(\br,\br_c)|\right]  \notag\\
        &\qquad\times|\Phi^{\alpha'}_{m,\bk''}(\br,\br_c)\rangle f^{\nu,\alpha'}_{m,\bk''} \left( f^{\nu,\beta'}_{m',\boldsymbol{k}''} \right)^*\langle  \Phi^{\beta'}_{m',\bk''}(\br,\br_c)|\hat{\boldsymbol{r}}|\Phi^{\beta}_{n',\bk'}(\br,\br_c)\rangle f^{\mu,\beta}_{n',\boldsymbol{k}'}e^{-i\gamma(\boldsymbol{k}',\tau_s)}|a(\boldsymbol{k}',\tau_s)|\Big\rangle\notag\\
        &= \Big\langle\hbar\sum_{nn',\alpha\beta}\sum_{\nu}\sum_{\alpha'\beta',mm'} \int d\boldsymbol{k}d\boldsymbol{k}'d\bk''\, |a(\boldsymbol{k},\tau_s)|e^{i\gamma(\boldsymbol{k},\tau_s)}[\omega_\mu(\bk,\br_c)-\omega_{\nu}(\bk'',\br_c)+(n-m)\Omega]\frac{\partial}{\partial\boldsymbol{r}_c}\left[\left( f^{\mu,\alpha}_{n,\boldsymbol{k}} \right)^*\langle  \Phi^\alpha_{n,\bk}(\br,\br_c)|\right]  \notag\\
        &\qquad\times|\Phi^{\alpha'}_{m,\bk''}(\br,\br_c)\rangle f^{\nu,\alpha'}_{m,\bk''} \left( f^{\nu,\beta'}_{m',\boldsymbol{k}''} \right)^* \Big\{i\partial_{\bk'}\delta_{\beta,\beta'}e^{-i(n'-m')\Omega \tau_f}\delta(\bk''_{m'}-\bk'_{n'})\notag\\
        &\qquad + \langle \Phi^{\beta'}_{m',\bk''}(\br,\br_c)|e^{i\bk'_{n'}\cdot\br}i\partial_{\bk'}\left[e^{-i\bk'_{n'}\cdot\br}|\Phi^{\beta}_{n',\bk'}(\br,\br_c)\rangle\right]\Big\}f^{\mu,\beta}_{n',\boldsymbol{k}'}e^{-i\gamma(\boldsymbol{k}',\tau_s)}|a(\boldsymbol{k}',\tau_s)| \Big\rangle,
\end{align}
where we have used the relation~\cite{blount1962formalisms}:
\begin{equation}
    \langle  \Phi^{\beta'}_{m',\bk''}(\br,\br_c)|\hat{\boldsymbol{r}}|\Phi^{\beta}_{n',\bk'}(\br,\br_c)\rangle=i\partial_{\bk'}\delta_{\beta,\beta'}e^{-i(n'-m')\Omega \tau_f}\delta(\bk''_{m'}-\bk'_{n'})+ \langle \Phi^{\beta'}_{m',\bk''}(\br,\br_c)|e^{i\bk'_{n'}\cdot\br}i\partial_{\bk'}\left[e^{-i\bk'_{n'}\cdot\br}|\Phi^{\beta}_{n',\bk'}(\br,\br_c)\rangle\right],
\end{equation}
which is an extension of Eq.~\eqref{positionInBloch}. Now, we have to be very careful in evaluating each term in the above expression. The term involving $i\partial_{\bk}\delta_{\beta,\beta'}e^{-i(n'-m')\Omega \tau_f} $ is
\begin{align}
        &\Big\langle\hbar\sum_{nn',\alpha\beta}\sum_{\nu}\sum_{\alpha'\beta',mm'} \int d\boldsymbol{k}d\boldsymbol{k}'d\bk''\, |a(\boldsymbol{k},\tau_s)|e^{i\gamma(\boldsymbol{k},\tau_s)}[\omega_\mu(\bk,\br_c)-\omega_{\nu}(\bk'',\br_c)+(n-m)\Omega]\frac{\partial}{\partial\boldsymbol{r}_c}\left[\left( f^{\mu,\alpha}_{n,\boldsymbol{k}} \right)^*\langle  \Phi^\alpha_{n,\bk}(\br,\br_c)|\right]  \notag\\
        &\qquad\times|\Phi^{\alpha'}_{m,\bk''}(\br,\br_c)\rangle f^{\nu,\alpha'}_{m,\bk''} \left( f^{\nu,\beta'}_{m',\boldsymbol{k}''} \right)^*i\partial_{\bk'}\delta_{\beta,\beta'}e^{-i(n'-m')\Omega \tau_f}\delta(\bk''_{m'}-\bk'_{n'})f^{\mu,\beta}_{n',\boldsymbol{k}'}e^{-i\gamma(\boldsymbol{k}',\tau_s)}|a(\boldsymbol{k}',\tau_s)| \Big\rangle \notag\\
        &=\Big\langle\hbar\sum_{nn',\alpha\beta}\sum_{\nu}\sum_{\alpha',ml} \int d\bk d\bk'\, |a(\boldsymbol{k},\tau_s)|e^{i\gamma(\boldsymbol{k},\tau_s)}[\omega_\mu(\bk,\br_c)-\omega_{\nu}(\bk'+l\boldsymbol{\kappa},\br_c)+(n-m)\Omega]\frac{\partial}{\partial\boldsymbol{r}_c}\left[\left( f^{\mu,\alpha}_{n,\boldsymbol{k}} \right)^*\langle  \Phi^\alpha_{n,\bk}(\br,\br_c)|\right]  \notag\\
        &\qquad\times|\Phi^{\alpha'}_{m,\bk'+l\boldsymbol\kappa}(\br,\br_c)\rangle f^{\nu,\alpha'}_{m,\bk'+l\boldsymbol\kappa} \left( f^{\nu,\beta}_{n'-l,\bk'+l\boldsymbol\kappa} \right)^*i\partial_{\bk'}e^{-il\Omega \tau_f}f^{\mu,\beta}_{n',\boldsymbol{k}'}e^{-i\gamma(\boldsymbol{k}',\tau_s)}|a(\boldsymbol{k}',\tau_s)| \Big\rangle \notag\\
        &=\Big\langle\hbar\sum_{n,\alpha}\sum_{\nu}\sum_{\alpha',ml} \int d\bk d\bk'\, |a(\boldsymbol{k},\tau_s)|e^{i\gamma(\boldsymbol{k},\tau_s)}[\omega_\mu(\bk,\br_c)-\omega_{\nu}(\bk'+l\boldsymbol{\kappa},\br_c)+(n-m)\Omega]\frac{\partial}{\partial\boldsymbol{r}_c}\left[\left( f^{\mu,\alpha}_{n,\boldsymbol{k}} \right)^*\langle  \Phi^\alpha_{n,\bk}(\br,\br_c)|\right]  \notag\\
        &\qquad\times|\Phi^{\alpha'}_{m,\bk'+l\boldsymbol\kappa}(\br,\br_c)\rangle f^{\nu,\alpha'}_{m,\bk'+l\boldsymbol\kappa} e^{-il\Omega \tau_f}\left\{\sum_{n',\beta}\left[\left( f^{\nu,\beta}_{n'-l,\bk'+l\boldsymbol\kappa} \right)^*f^{\mu,\beta}_{n',\boldsymbol{k}'}\right]i\partial_{\bk'}e^{-i\gamma(\boldsymbol{k}',\tau_s)}|a(\boldsymbol{k}',\tau_s)| \right.\notag\\
        &\qquad+ \left.e^{-i\gamma(\boldsymbol{k}',\tau_s)}|a(\boldsymbol{k}',\tau_s)| \sum_{n',\beta}\left[\left( f^{\nu,\beta}_{n'-l,\bk'+l\boldsymbol\kappa} \right)^*i\partial_{\bk'}f^{\mu,\beta}_{n',\boldsymbol{k}'}\right]\right\}\Big\rangle\notag\\
        &=\Big\langle\hbar\sum_{nn',\alpha\beta}\sum_{\nu}\sum_{\alpha',ml} \int d\bk d\bk'\, |a(\boldsymbol{k},\tau_s)|e^{i\gamma(\boldsymbol{k},\tau_s)}[\omega_\mu(\bk,\br_c)-\omega_{\nu}(\bk'+l\boldsymbol{\kappa},\br_c)+(n-m)\Omega]\frac{\partial}{\partial\boldsymbol{r}_c}\left[\left( f^{\mu,\alpha}_{n,\boldsymbol{k}} \right)^*\langle  \Phi^\alpha_{n,\bk}(\br,\br_c)|\right]  \notag\\
        &\qquad\times|\Phi^{\alpha'}_{m,\bk'+l\boldsymbol\kappa}(\br,\br_c)\rangle f^{\nu,\alpha'}_{m,\bk'+l\boldsymbol\kappa} e^{-il\Omega \tau_f}e^{-i\gamma(\boldsymbol{k}',\tau_s)}|a(\boldsymbol{k}',\tau_s)|\left( f^{\nu,\beta}_{n'-l,\bk'+l\boldsymbol\kappa} \right)^*i\partial_{\bk'}f^{\mu,\beta}_{n',\boldsymbol{k}'}\Big\rangle,
\end{align}
where we have used Eq.~\eqref{orthogonality} and Eq.~\eqref{MomentumConservation}.
Plugging the above result into Eq.~\eqref{secondTerm}, the second term becomes
\begin{align}
        &\Big\langle\hbar\sum_{nn',\alpha\beta}\sum_{\nu}\sum_{\alpha'\beta',mm'} \int d\boldsymbol{k}d\boldsymbol{k}'d\bk''\, |a(\boldsymbol{k},\tau_s)|e^{i\gamma(\boldsymbol{k},\tau_s)}e^{-i\gamma(\boldsymbol{k}',\tau_s)}|a(\boldsymbol{k}',\tau_s)|  \notag\\
        &\qquad\times[\omega_\mu(\bk,\br_c)-\omega_{\nu}(\bk'',\br_c)+(n-m)\Omega]\frac{\partial}{\partial\boldsymbol{r}_c}\left[\left( f^{\mu,\alpha}_{n,\boldsymbol{k}} \right)^*\langle  \Phi^\alpha_{n,\bk}(\br,\br_c)|\right] |\Phi^{\alpha'}_{m,\bk''}(\br,\br_c)\rangle f^{\nu,\alpha'}_{m,\bk''} \left( f^{\nu,\beta'}_{m',\boldsymbol{k}''} \right)^*\notag\\
        &\qquad\times 
        \langle \Phi^{\beta'}_{m',\bk''}(\br,\br_c)|e^{i\bk'_{n'}\cdot\br}i\partial_{\bk'}\left[e^{-i\bk'_{n'}\cdot\br}f^{\mu,\beta}_{n',\boldsymbol{k}'}|\Phi^{\beta}_{n',\bk'}(\br,\br_c)\rangle\right]\Big\rangle\notag\\
        &= \Big\langle\sum_{nn',\alpha\beta}\sum_{\nu}\sum_{\alpha'\beta',mm'} \int d\boldsymbol{k}d\boldsymbol{k}'d\bk''\, |a(\boldsymbol{k},\tau_s)|e^{i\gamma(\boldsymbol{k},\tau_s)}e^{-i\gamma(\boldsymbol{k}',\tau_s)}|a(\boldsymbol{k}',\tau_s)|  \notag\\
        &\qquad\times\frac{\partial}{\partial\boldsymbol{r}_c}\left[\left( f^{\mu,\alpha}_{n,\boldsymbol{k}} \right)^*\langle  \Phi^\alpha_{n,\bk}(\br,\br_c)|\right][\hbar\omega_\mu(\bk,\br_c)+n\hbar\Omega-\hat{H}(\br,\br_c)] |\Phi^{\alpha'}_{m,\bk''}(\br,\br_c)\rangle f^{\nu,\alpha'}_{m,\bk''} \left( f^{\nu,\beta'}_{m',\boldsymbol{k}''} \right)^*\notag\\
        &\qquad\times 
        \langle \Phi^{\beta'}_{m',\bk''}(\br,\br_c)|e^{i\bk'_{n'}\cdot\br}i\partial_{\bk'}\left[e^{-i\bk'_{n'}\cdot\br}f^{\mu,\beta}_{n',\boldsymbol{k}'}|\Phi^{\beta}_{n',\bk'}(\br,\br_c)\rangle\right]\Big\rangle\notag\\
        & = \Big\langle\sum_{nn',\alpha\beta} \int d\bk d\bk' \, |a(\boldsymbol{k},\tau_s)|e^{i\gamma(\boldsymbol{k},\tau_s)}e^{-i\gamma(\boldsymbol{k}',\tau_s)}|a(\boldsymbol{k}',\tau_s)|\frac{\partial}{\partial\boldsymbol{r}_c}\left[e^{i\bk_n\cdot\br}\left( f^{\mu,\alpha}_{n,\boldsymbol{k}} \right)^*\langle  \Phi^\alpha_{n,\bk}(\br,\br_c)|\right]  \notag\\
        &\qquad\times e^{-i\bk_n\cdot\br}[\hbar\omega_\mu(\bk,\br_c)+n\hbar\Omega-\hat{H}(\br,\br_c)] 
        e^{i\bk'_{n'}\cdot\br}i\partial_{\bk'}\left[e^{-i\bk'_{n'}\cdot\br}f^{\mu,\beta}_{n',\boldsymbol{k}'}|\Phi^{\beta}_{n',\bk'}(\br,\br_c)\rangle\right]\Big\rangle.
\end{align}
\end{widetext}
After taking the time average, we will have $\delta_{n,n'}$ and approximately momentum conservation $\delta(\bk-\bk')$, which gives
\begin{align}
        &\Big\langle\sum_{nn',\alpha\beta} \int d\bk  \, |a(\boldsymbol{k},\tau_s)|^2\frac{\partial}{\partial\boldsymbol{r}_c}\left[e^{i\bk\cdot\br}\left( f^{\mu,\alpha}_{n,\boldsymbol{k}} \right)^*\langle  \Phi^\alpha_{n,\bk}(\br,\br_c)|\right]  \notag\\
        &\qquad\times e^{-i\bk\cdot\br}[\hbar\omega_\mu(\bk,\br_c)+n\hbar\Omega-\hat{H}(\br,\br_c)] 
        e^{i\bk\cdot\br}\notag\\
        &\qquad\times i\partial_{\bk}\left[e^{-i\bk\cdot\br}f^{\mu,\beta}_{n',\bk}|\Phi^{\beta}_{n',\bk}(\br,\br_c)\rangle\right]\Big\rangle \notag\\
        &= \Big\langle \int d\bk  \, |a(\boldsymbol{k},\tau_s)|^2\frac{\partial}{\partial\boldsymbol{r}_c}\sum_{n,\alpha}\left[e^{i\bk\cdot\br}\left( f^{\mu,\alpha}_{n,\boldsymbol{k}} \right)^*\langle  \Phi^\alpha_{n,\bk}(\br,\br_c)|\right]  \notag\\
        &\qquad\times e^{-i\bk\cdot\br}[\hbar\omega_\mu(\bk,\br_c)+i\hbar\partial_{\tau_f}-\hat{H}(\br,\br_c)] 
        e^{i\bk\cdot\br}\notag\\
        &\qquad\times i\partial_{\bk}\sum_{n',\beta}\left[e^{-i\bk\cdot\br}f^{\mu,\beta}_{n',\bk}|\Phi^{\beta}_{n',\bk}(\br,\br_c)\rangle\right]\Big\rangle 
\end{align}
Recall that the periodic part of the Floquet-Bloch wavefunction is defined as
\begin{equation}
    | \Tilde{u}_{\boldsymbol{k}}^{\mu} \rangle = \sum_{n,\alpha} f^{\mu,\alpha}_{n,\bk}e^{-in\Omega \tau_f}e^{in\boldsymbol{\kappa}\cdot\br}|u^\alpha_{n,\bk}\rangle = e^{-i\bk\cdot\br}\sum_{n,\alpha}f^{\mu,\alpha}_{n,\bk}|\Phi^{\alpha}_{n,\bk}\rangle
\end{equation}
which satisfies the Schrodinger equation at specific $\bk$:
\begin{equation}
    [\hat{H}(\bk)-i\hbar\partial_{\tau_f}]| \Tilde{u}_{\boldsymbol{k}}^{\mu} \rangle = \hbar\omega_\mu(\bk) | \Tilde{u}_{\boldsymbol{k}}^{\mu} \rangle
\end{equation}
with $\hat{H}(\bk)\equiv e^{-i\bk\cdot\br}\hat{H}(\br)e^{i\bk\cdot\br} $ and also the completeness relation: $\sum_\mu | \Tilde{u}_{\boldsymbol{k}}^{\mu} \rangle\langle\Tilde{u}_{\boldsymbol{k}}^{\mu}| = \mathcal{I}_\bk$. Thus, the second term is reduced to
\begin{equation}
    \begin{split}
\langle\langle \partial_{\br_c}\Tilde{u}_{\bk_c}^{\mu}|\cdot\left[\hbar\omega_\mu(\bk_c,\br_c)-\left(\hat{H}(\bk_c,\br_c)-i\hbar\partial_{\tau_f}\right)\right]|i\partial_{\bk_c}\Tilde{u}_{\bk_c}^{\mu}\rangle\rangle
    \end{split}
\end{equation}
Finally, we obtain the energy correction as
\begin{align}
       & \Delta\mathcal{E}_\mu = \Re\,\langle\langle W_\mu | \frac{\partial\hat{H}_c}{\partial\boldsymbol{r}_c}\cdot(\hat{\boldsymbol{r}}-\boldsymbol{r}_c)| W_\mu \rangle\rangle \notag\\
        &= \Re\left\{ \frac{\partial\hbar\omega_\mu(\bk_c,\br_c)}{\partial\br_c}\cdot\br_c + \langle\langle \frac{\partial}{\partial {\br_c}}\Tilde{u}_{\bk_c}^{\mu}|\right.\notag\\
        &\quad\cdot\left[\hbar\omega_\mu(\bk_c,\br_c)-\left(\hat{H}(\bk_c,\br_c)-i\hbar\frac{\partial}{\partial \tau_f}\right)\right]|i\frac{\partial}{\partial\bk_c}\Tilde{u}_{\bk_c}^{\mu}\rangle\rangle \notag\\
        &\quad- \left.\frac{\partial\hbar\omega_\mu(\bk_c,\br_c)}{\partial\br_c}\cdot\br_c  \right\} \notag\\
        &= \Im \langle\langle \frac{\partial}{\partial {\br_c}}\Tilde{u}_{\bk_c}^{\mu}|\notag\\
        &\quad\cdot\left[\left(\hat{H}(\bk_c,\br_c)-i\hbar\frac{\partial}{\partial \tau_f}\right)-\hbar\omega_\mu(\bk_c,\br_c)\right]|\frac{\partial}{\partial\bk_c}\Tilde{u}_{\bk_c}^{\mu}\rangle\rangle \notag\\
        &= \Im \langle\langle \frac{\partial}{\partial {\br_c}}\Tilde{u}_{\bk_c}^{\mu}|\cdot\left[\hat{H}^F(\bk_c,\br_c)-\hbar\omega_\mu(\bk_c,\br_c)\right]|\frac{\partial}{\partial\bk_c}\Tilde{u}_{\bk_c}^{\mu}\rangle\rangle,
\end{align}
where $\hat{H}^F = \hat{H}-i\hbar\partial_{\tau_f}$ is the Floquet Hamiltonian. Here, $\partial_{\tau_f}$ only acts on the fast-varying time-dependent variables, which is different from $\partial_{\tau_s}$ as we explained before. In the case where there is no external inhomogeneity in time (i.e., no slow time dependence), we may write $\hat{H}^F = \hat{H}-i\hbar\partial_{t}$ as shown in Ref.~\cite{topp2022orbital}.

In Sec.~\ref{EM_response}, we specified the external slowly varying field as the electromagnetic field, thus the energy correction can be written as
\begin{align}
        &\Delta\mathcal{E}_\mu = \Im \langle\langle \frac{\partial}{\partial {\br_c}}\Tilde{u}_{\bk_c}^{\mu}(\br_c,\tau_s)|\notag\\
        &\quad\cdot\left[\hat{H}^F(\bk_c,\br_c)-\hbar\omega_\mu(\bk_c,\br_c)\right]|\frac{\partial}{\partial\bk_c}\Tilde{u}_{\bk_c}^{\mu}(\br_c,\tau_s)\rangle\rangle \notag\\
        &= \Im \left[\frac{\partial e\bA}{\hbar\partial \br_c}\langle\langle \frac{\partial}{\partial {\bq_c}}\Tilde{u}_{\bq_c}^{\mu}|\right]\cdot\left[\hat{H}^F(\bq_c)-\hbar\omega_\mu(\bq_c)\right]|\frac{\partial}{\partial\bq_c}\Tilde{u}_{\bq_c}^{\mu}\rangle\rangle \notag\\
        &= -\frac{e}{2\hbar}(\boldsymbol\nabla_{\br_c}\times\bA)\notag\\
        &\quad\cdot\Im\langle\langle \frac{\partial}{\partial {\bq_c}}\Tilde{u}_{\bq_c}^{\mu}|\times\left[\hat{H}^F(\bq_c)-\hbar\omega_\mu(\bq_c)\right]|\frac{\partial}{\partial\bq_c}\Tilde{u}_{\bq_c}^{\mu}\rangle\rangle\notag\\
        &\equiv  -\bB\cdot\bm,
\end{align}
where $ \bB = \boldsymbol\nabla_{\br_c}\times\bA(\br_c,\tau_s)$ is the magnetic field and 
\begin{equation}
    \bm = \frac{e}{2\hbar}\Im\langle\langle \frac{\partial}{\partial {\bq_c}}\Tilde{u}_{\bq_c}^{\mu}|\times\left[\hat{H}^F(\bq_c)-\hbar\omega_\mu(\bq_c)\right]|\frac{\partial}{\partial\bq_c}\Tilde{u}_{\bq_c}^{\mu}\rangle\rangle
\end{equation}
is the local magnetic moment of the Floquet-Bloch wave-packet due to its self-rotation in the magnetic field.

\section{The integration and summation in Eq.~\eqref{DensityOperator_OSTC}}\label{change_integration}
The integration over the quasi-momentum $\bk$ and summation over the index $n$ (and $m$) should make the following equation true:
\begin{equation}
\begin{split}
    &\sum_{n}\int' d\bk |\Psi^{\mu}_{\bk+n\boldsymbol{\kappa}}(\br,t)\rangle\langle\Psi^\nu_{\bk+n\boldsymbol{\kappa}}(\br,t)| \\
    &= \int_{BZ} d\bk |\Psi^{\mu}_{\bk}(\br,t)\rangle\langle\Psi^\nu_{\bk}(\br,t)|,
\end{split}
\end{equation}
so that it can run through all possible states in the first Floquet-Bloch Brillouin zone. There are two possible scenarios: commensurate and incommensurate pair of $(\boldsymbol{\kappa},\bG)$, where $\bG$ is the reciprocal lattice vector. Let us use a simple (1+1)D space-time crystal as an example. For the commensurate case, we have $\kappa=pG/q$ with ($p,q$) being co-prime numbers, where we should have
\begin{equation}
    \sum_{n}\int' dk \to \sum_{n=0}^{q-1}\int_{0}^{G/q} dk.
\end{equation}
For the incommensurate case, one should expect that $q\to \infty$, which reduces the above mapping to simply $(1/\mathcal{N})\sum_{n=0}^{\infty}$ with no integration over $k$ needed (here $\mathcal{N}$ is a proper normalization factor), instead, we need to choose a representative $k_0$ for the quasi-momentum. This is counterintuitive but can be understood in the following way: the set $\{k_0+n\kappa\mod G|n\in\mathbb{N}\}$ is a dense cover of the interval $[0,G]$ if $\kappa$ and $G$ are incommensurate to each other. 

\section{Density matrix of the Floquet-Bloch system with damping}\label{densityMatrixFordamping}
Casting Eq.~\eqref{LiouvilleEq} into the Floquet-Bloch basis, we have
\begin{align}
&\langle\Psi^\mu_{\bk+n\boldsymbol{\kappa}}(\br,t)|i\partial_{t}\hat{\rho}|\Psi^\nu_{\bk+m\boldsymbol{\kappa}}(\br,t)\rangle \notag \\
&= i\partial_{t}\langle\Psi^\mu_{\bk+n\boldsymbol{\kappa}}(\br,t)|\hat{\rho}|\Psi^\nu_{\bk+m\boldsymbol{\kappa}}(\br,t)\rangle \notag\\
&\quad- \langle i\partial_{t}\Psi^\mu_{\bk+n\boldsymbol{\kappa}}(\br,t)|\hat{\rho}|\Psi^\nu_{\bk+m\boldsymbol{\kappa}}(\br,t)\rangle \notag\\
&\quad-\langle\Psi^\mu_{\bk+n\boldsymbol{\kappa}}(\br,t)|\hat{\rho}|i\partial_{t}\Psi^\nu_{\bk+m\boldsymbol{\kappa}}(\br,t)\rangle \notag\\
&= i\partial_{t} \rho^{F}_{\{\mu,n\},\{\nu,m\}}(\bk,t) + \langle\Psi^\mu_\bk(\br,t)|[\hat{H},\hat{\rho}]|\Psi^\nu_\bk(\br,t)\rangle \notag\\
&\quad- [\omega_\mu(\bk+n\boldsymbol{\kappa})-\omega_\nu(\bk+m\boldsymbol{\kappa})]\rho^{F}_{\mu,\nu}(\bk,t),
\end{align}
and
\begin{align}
        &\langle\Psi^\mu_{\bk+n\boldsymbol{\kappa}}(\br,t)|[\hat{D},\hat{\rho}]|\Psi^\nu_{\bk+m\boldsymbol{\kappa}}(\br,t)\rangle \notag\\
        &= -\Gamma \langle\Psi^\mu_{\bk+n\boldsymbol{\kappa}}(\br,t)|\hat{\rho}|\Psi^\nu_{\bk+m\boldsymbol{\kappa}}(\br,t)\rangle \notag\\
        &\quad+\Gamma \langle\Psi^\mu_{\bk+n\boldsymbol{\kappa}}(\br,t)|\hat{\rho}^{B,eq}|\Psi^\nu_{\bk+m\boldsymbol{\kappa}}(\br,t)\rangle \notag\\
        &= -\Gamma \rho^{F}_{\{\mu,n\},\{\nu,m\}}(\bk,t) \notag\\
        &\quad+ \Gamma \sum_{\alpha\beta,p q}\rho^{B,eq}_{\alpha,\beta}(\bk+n\boldsymbol{\kappa}+p\boldsymbol\kappa,\bk+m\boldsymbol{\kappa}+q\boldsymbol{\kappa})\notag\\
        &\qquad\times e^{i(p-q)\Omega t}\left( f^{\mu,\alpha}_{p,\bk+n\boldsymbol{\kappa}} \right)^*f^{\nu,\beta}_{q,\bk+m\boldsymbol{\kappa}}\notag\\
        &= -\Gamma \rho^{F}_{\{\mu,n\},\{\nu,m\}}(\bk,t) \notag\\
        &\quad+ \Gamma \sum_{\alpha\beta,p l}\rho^{B,eq}_{\alpha,\beta}(\bk+n\boldsymbol{\kappa}+p\boldsymbol\kappa)\delta_{n-m,l}\notag\\
        &\qquad\times e^{-i l\Omega t}\left( f^{\mu,\alpha}_{p,\bk+n\boldsymbol{\kappa}} \right)^*f^{\nu,\beta}_{p+l,\bk+m\boldsymbol{\kappa}},
\end{align}
where $l \equiv q-p$ and we have assumed that the Bloch equilibrium is diagonal in momentum $\bk$: $\rho^{B,eq}_{\alpha,\beta}(\bk,\bk') = \rho^{B,eq}_{\alpha,\beta}(\bk) \delta(\bk-\bk')$.
Putting all together, the Liouville equation can be rewritten as
\begin{align}
    &[i\partial_{t}-\omega_\mu(\bk+n\boldsymbol{\kappa})+\omega_\nu(\bk+m\boldsymbol{\kappa})+i\Gamma]\rho^{F}_{\{\mu,n\},\{\nu,m\}}(\bk,t)\notag\\
    &= i\Gamma \sum_{\alpha\beta,p l}\rho^{B,eq}_{\alpha,\beta}(\bk+n\boldsymbol{\kappa}+p\boldsymbol\kappa)\delta_{n-m,l}\notag\\
    &\qquad\times e^{-i l\Omega t}\left( f^{\mu,\alpha}_{p,\bk+n\boldsymbol{\kappa}} \right)^*f^{\nu,\beta}_{p+l,\bk+m\boldsymbol{\kappa}}.
\end{align}
Since we are considering the steady states, the density matrix must also be time-periodic, so that we can write
\begin{equation}
    \rho^{F}_{\{\mu,n\},\{\nu,m\}}(\bk,t) = \sum_l \rho^{F,l}_{\{\mu,n\},\{\nu,m\}}(\bk)e^{-il\Omega t}.
\end{equation}
we can obtain:
\begin{align}\label{densityMatrix_appendix}
    &\rho^{F,l}_{\{\mu,n\},\{\nu,m\}}(\bk) =-i\Gamma\sum_{\alpha\beta,p}\delta_{n-m,l}\notag\\
    &\quad\times\frac{\rho^{B,eq}_{\alpha,\beta}(\bk+n\boldsymbol{\kappa}+p\boldsymbol\kappa)\left( f^{\mu,\alpha}_{p,\bk+n\boldsymbol{\kappa}} \right)^*f^{\nu,\beta}_{p+l,\bk+m\boldsymbol{\kappa}}}{\omega_\mu(\bk+n\boldsymbol{\kappa})-n\Omega-\omega_\nu(\bk+m\boldsymbol{\kappa})+m\Omega-i\Gamma}\notag\\
    &=i\Gamma\sum_{\alpha\beta,p} \frac{\rho^{B,eq}_{\alpha,\beta}(\bk+n\boldsymbol{\kappa}+p\boldsymbol\kappa)\left( f^{\mu,\alpha}_{p,\bk+n\boldsymbol{\kappa}} \right)^*f^{\nu,\beta}_{p+l,\bk+m\boldsymbol{\kappa}}}{\omega^n_\mu(\bk)-\omega^m_\nu(\bk)-i\Gamma}\delta_{n-m,l}.
\end{align}
Here $\omega_\mu^n(\bk)\equiv\omega_\mu(\bk+l\boldsymbol{\kappa})-n\Omega$ is actually corresponding to the $\mu$-th Floquet-Bloch band at the $n$-th Floquet replica. One can think about this as an extended zero picture in the frequency domain~\cite{gao2021floquet}.

\section{Density matrix for $\kappa=0$ with finite damping}\label{densitymatrixrectangular}
Our discussion now will be restricted to the case where $\boldsymbol{\kappa}=0$, then the density matrix becomes~\cite{gao2022dc}:
\begin{equation}\label{densityMatrix_RSTC}
    \rho^{F,l}_{\mu,\nu}(\bk) = -i\Gamma\sum_{\alpha\beta,n}\frac{\rho^{B,eq}_{\alpha,\beta}(\bk)\left( f^{\mu,\alpha}_{n,\bk} \right)^*f^{\nu,\beta}_{n+l,\bk}}{\omega_\mu(\bk)-\omega_\nu(\bk)-l\Omega-i\Gamma}.
\end{equation}
which is clearly non-diagonal and time-dependent for finite damping. For now on, to simplify our discussion, we consider only the case where the damping is very small such that $\Gamma\ll \Delta^{FB}$. Here $\Delta^{FB}=\min_{\mu,\nu,l,\bk}(\omega_\mu(\bk)-\omega_\nu(\bk)-l\Omega)$ is the Floquet-Bloch direct band gap. Our first goal is to diagonalize the density matrix in Eq.~\eqref{densityMatrix_RSTC} up to the first order in $\Gamma$, which can be done by properly rotating the non-damped basis $\{ |\Psi^\mu_\bk(\br,t)\rangle \}$:
\begin{align}
    &|\Psi^\mu_\bk(\br,t;\Gamma)\rangle \notag\\
    &= |\Psi^\mu_\bk(\br,t)\rangle + \Gamma\sum_{\nu}\mathcal{G}_{\mu,\nu}(\bk,t)|\Psi^\nu_\bk(\br,t)\rangle + O(\Gamma^2).
\end{align}
One should ensure that the new basis $\{ |\Psi^\mu_\bk(\br,t;\Gamma)\rangle \}$ has to be also orthonormalized, namely,
\begin{equation}
\begin{split}
    &\langle \Psi^\mu_\bk(\br,t;\Gamma)|\Psi^\nu_{\bk'}(\br,t;\Gamma)\rangle \\
    &= [\delta_{\mu,\nu}+\Gamma(\mathcal{G}_{\nu,\mu}(\bk,t)+\mathcal{G}^*_{\mu,\nu}(\bk,t))]\delta(\bk-\bk')+ O(\Gamma^2) \\
    &= \delta_{\mu,\nu}\delta(\bk-\bk')+ O(\Gamma^2),
\end{split}
\end{equation}
which requires that
\begin{equation}\label{G_relation}
    \mathcal{G}_{\nu,\mu}(\bk,t)=-\mathcal{G}^*_{\mu,\nu}(\bk,t).
\end{equation}
Then the density matrix under this new basis becomes
\begin{align}
        &\rho^F_{\mu,\nu}(\bk,t;\Gamma) = \langle \Psi^\mu_\bk(\br,t;\Gamma)|\hat{\rho}|\Psi^\nu_{\bk}(\br,t;\Gamma)\rangle= \rho^F_{\mu,\mu}(\bk)\delta_{\mu,\nu} \notag\\
        & \quad-i\Gamma\sum_{\alpha\beta,n}\sum_{l(\mu,\nu)}\frac{\rho^{B,eq}_{\alpha,\beta}(\bk)\left( f^{\mu,\alpha}_{n,\bk} \right)^*f^{\nu,\beta}_{n+l,\bk}}{\omega_\mu(\bk)-\omega_\nu(\bk)-l\Omega-i\Gamma}e^{-il\Omega t} \notag\\
        &\quad+ \Gamma\mathcal{G}_{\nu,\mu}(\bk,t)\left[\rho^F_{\mu,\mu}(\bk)-\rho^F_{\nu,\nu}(\bk)\right] + O(\Gamma^2),
\end{align}
where
\begin{equation}
    \rho^F_{\mu,\mu}(\bk) = \sum_{\alpha\beta,n}\rho^{B,eq}_{\alpha,\beta}(\bk)\left( f^{\mu,\alpha}_{n,\bk} \right)^*f^{\mu,\beta}_{n,\bk}
\end{equation}
is the time-independent and diagonal density matrix in Eq.~\eqref{densityMatrix_RSTC} when $\Gamma\to 0$, $\sum_{l(\mu,\nu)}$ stands for a conditional summation over index $l$: summing over all $l$ if $\mu\neq\nu$ or summing over non-zero $l$ if $\mu=\nu$. Thus, for $\rho^F_{\mu,\nu}(\bk,t;\Gamma)$ to be diagonalized up to the first order in $\Gamma$, we have, for $\mu\neq\nu $
\begin{align}
    &\mathcal{G}_{\nu,\mu}(\bk,t) = \frac{i}{\rho^F_{\mu,\mu}(\bk)-\rho^F_{\nu,\nu}(\bk)}\notag\\
    &\times\sum_{\alpha\beta,n}\sum_{l}\frac{\rho^{B,eq}_{\alpha,\beta}(\bk)\left( f^{\mu,\alpha}_{n,\bk} \right)^*f^{\nu,\beta}_{n+l,\bk}}{\omega_\mu(\bk)-\omega_\nu(\bk)-l\Omega-i\Gamma}e^{-il\Omega t}.
\end{align}
One can check that Eq.~\eqref{G_relation} indeed holds.
However, the diagonal terms in $\mathcal{G}$ can not be determined which are only known to be purely imaginary. The final diagonalized density matrix, therefore, reads
\begin{align}
    &\rho^F_{\mu,\nu}(\bk,t;\Gamma) = \delta_{\mu,\nu}\times\notag\\
    &\left[ \rho^F_{\mu,\mu}(\bk) + i\Gamma\sum_{\alpha\beta,n}\sum_{l\neq 0}\frac{\rho^{B,eq}_{\alpha,\beta}(\bk)\left( f^{\mu,\alpha}_{n,\bk} \right)^*f^{\mu,\beta}_{n+l,\bk}}{l\Omega+i\Gamma}e^{-il\Omega t} \right],
\end{align}
which will always contain fast-time-dependent parts. The discussion can be also applied to cases with not-so-small damping. In general, there is no such orthonormal basis that can make the density matrix diagonal and time-independent simultaneously for a non-zero $\Gamma$.

\begin{figure}[t]
    \centering
    \includegraphics[width=0.4\textwidth]{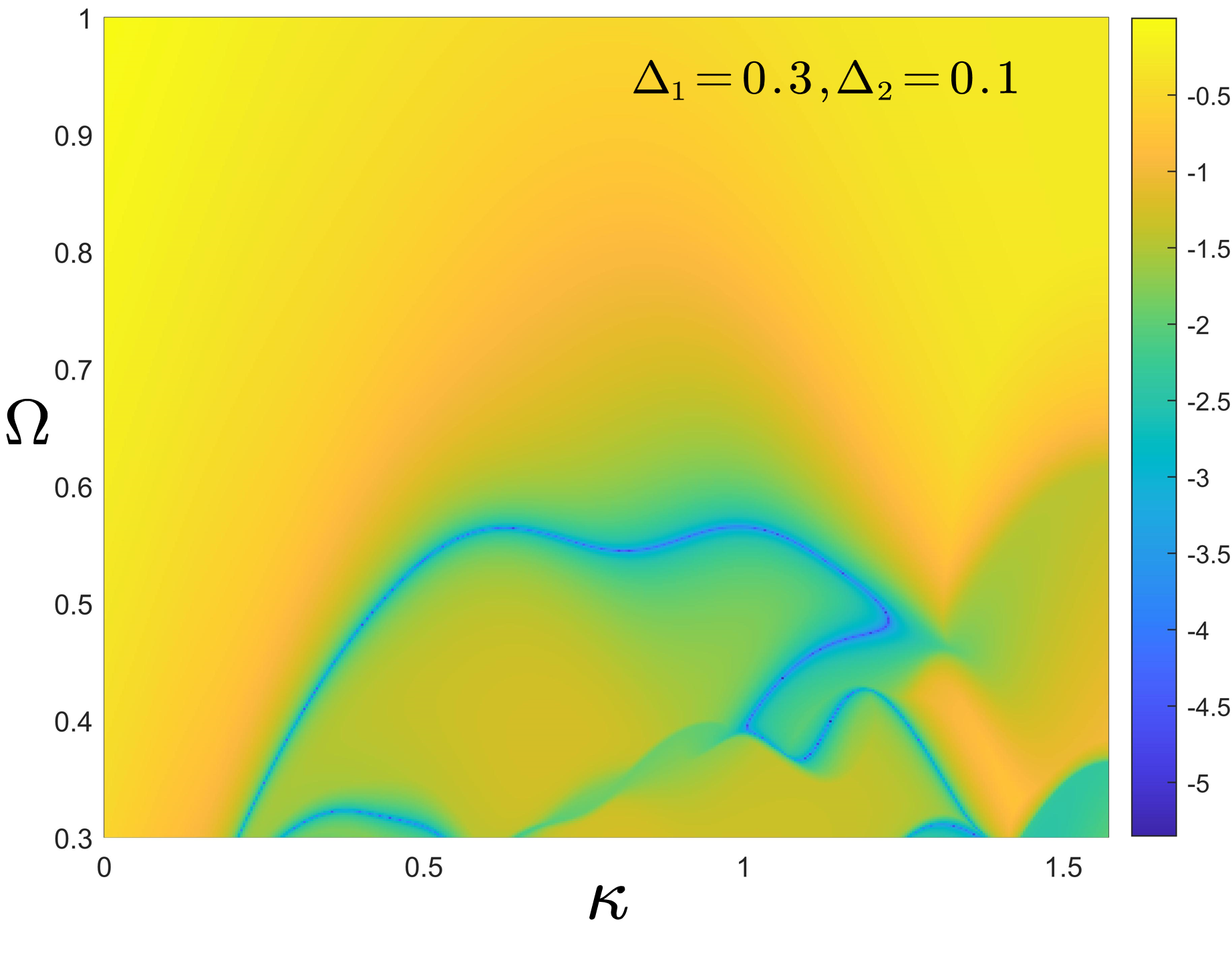}
    \caption{The Floquet direct gap (in logarithm $\log_{10}(\Delta^{FB}))$) of the model described in Eq.~\eqref{caseStudyHamiltonian} as a function of parameters ($\Omega,\kappa$) at fixed coupling strengths.}
    \label{Gap_plot}
\end{figure}
\section{Floquet gap of the oblique space-time crystal in ($\Omega,\kappa$) parameter space}\label{Floquet_gap}
In this Appendix, we show how the Floquet-Bloch direct band gap changes in the parameter space ($\Omega,\kappa$). The Floquet-Bloch direct band gap is defined as $\Delta^{FB}=\min_{\mu,\nu,n,m,\bk}(\omega^n_\mu(\bk)-\omega^m_\nu(\bk))$, where $\omega^n_\mu(\bk)\equiv\omega_\mu(\bk+n\boldsymbol{\kappa})-n\Omega$. In the case where only one Floquet-Bloch band is present, the gap is simply $\Delta^{FB}=\min_{\bk}(\omega(\bk)-\omega(\bk+\boldsymbol{\kappa})+\Omega)$.  As one can see clearly in Fig.~\ref{Gap_plot}, the Floquet gap shows a very similar pattern to the intrinsic current behavior discussed in the main text. Moreover, we also observe band-closing curves in the two-dimensional parameter space. It can be shown that the exact band-closing curves are the consequence of choosing $\Delta_1$ and $\Delta_2$ to be real, which makes the Floquet Hamiltonian in Eq.~\eqref{caseStudyHamiltonian} a real symmetric matrix.

\bibliographystyle{unsrt}

\begin{thebibliography}{10}
\bibitem{Note_kappa}
Here, we intentionally choose a specific $\kappa=2\pi/\sqrt{2}R$ that is irrationally related to the BZ size $2\pi/R$, so that no supercell in space can be chosen to reduce the oblique space-time crystal to a rectangular one.
\end{thebibliography}

\end{document}